\documentclass[11pt,a4paper]{article}

\usepackage{amsmath}
\usepackage{amssymb}
\usepackage{graphicx} 
\usepackage{hyperref} 
\usepackage[numbers,sort&compress]{natbib} 
\usepackage{booktabs} 
\usepackage{geometry} 
\usepackage{authblk} 
\usepackage{chemformula}
\usepackage{xcolor}
\usepackage{threeparttable}
\usepackage{caption} 
\usepackage{subcaption}
\usepackage{booktabs} 
\usepackage{multicol}
\usepackage{multirow}
\usepackage{sistyle}
\SIthousandsep{,}
\title{\textbf{Extending the Handover-Iterative VQE to Challenging Strongly Correlated Systems: \ch{N2} and Fe-S Cluster}} %

\author[1,a]{Pilsun Yoo}
\author[1]{Kyungmin Kim}
\author[1]{Eyuel E. Elala}
\author[1]{Shane McFarthing}
\author[1]{Aidan Pellow}
\author[1]{Johanna I. Fuks}
\author[1]{Doo Hyung Kang}
\author[1]{Pratanphorn Nakliang}
\author[1]{Jaewan Kim}
\author[2,a]{Himadri Pathak}
\author[2]{Tomonori Shirakawa}
\author[2]{Seiji Yunoki}
\author[1,*]{June-Koo Kevin Rhee}
\affil[1]{Qunova Computing, Inc., Daejeon, Rep. of Korea}
\affil[2]{RIKEN, Kobe, Japan}
\affil[a]{co-first author}
\date{\today} 

\begin{document}

\maketitle
\begin{abstract}
Accurately describing strongly correlated electronic systems remains a central challenge in quantum chemistry, as electron–electron interactions give rise to complex many-body wavefunctions that are difficult to capture with conventional approximations. Classical wavefunction-based approaches, such as the Semistochastic Heat-bath Configuration Interaction (SHCI) and the Density Matrix Renormalization Group (DMRG), currently define the state of the art, systematically converging toward the Full Configuration Interaction (FCI) limit, but at a rapidly increasing computational cost. Quantum computing algorithms promise to alleviate this scaling bottleneck by leveraging entanglement and superposition to represent correlated states more compactly. We introduced the Handover-Iterative Variational Quantum Eigensolver (HI-VQE) as a practical quantum computing algorithm with an iterative “handover” mechanism that dynamically exchanges information between quantum and classical computers, even using Noisy Intermediate-Scale Quantum (NISQ) computers. In this work, we extend the HI-VQE to benchmark two prototypical strongly correlated systems, the nitrogen molecule (\ch{N2}) and iron–sulfur (Fe–S) cluster, which serve as stringent tests for both classical and quantum electronic-structure methods. By comparing HI-VQE results against Heat-bath Configuration Interaction (HCI) benchmarks, we assess its accuracy, scalability, and ability to capture multireference correlation effects. Achieving quantitative agreement on these canonical systems demonstrates a viable pathway toward quantum-enhanced simulations of complex bioinorganic molecules, catalytic mechanisms, and correlated materials.
\end{abstract}

\section{Introduction}
Computing the ground-state energy of a many-electron system by solving the Schr\"odinger equation is the central task of quantum chemistry, and its cost grows exponentially with the number of correlated electrons and orbitals involved. Exact diagonalization in the Full Configuration Interaction (FCI) limit remains tractable only for a few tens of electrons, even as continuing advances in high-performance computing, parallelization, and GPU acceleration push this boundary outward\cite{Helgaker2000, Gao2024}. Beyond this scale, one must resort to approximate wavefunction- or density-based methods\cite{Cao2019Chemrev-QC,Friesner2005}.

Density Functional Theory (DFT) offers a favorable balance between accuracy and cost and scales to systems of thousands of atoms\cite{LargeHybridDFT2024}, but its single-particle construction breaks down the accurate description of strongly correlated systems\cite{Harvey2006_DFT_TransitionMetals,Cao2019_Nitrogenase_DFT,Vysotskiy2020_Nitrogenase_Model}. Coupled Cluster with Singles, Doubles, and perturbative Triples [CCSD(T)] is the single-reference gold standard, yet it too fails once static correlation dominates and the ground state can no longer be built from one Slater determinant\cite{bulik2015ccsd, Kowalski2000_CCSDT}. 

Selected configuration interaction (SCI) narrows this gap by identifying, rather than enumerating, the determinants that carry appreciable weight. Beginning with CIPSI\cite{huron1973iterative}, these methods grow a compact variational space iteratively and recover the remainder perturbatively, so that the selection operates within a chosen orbital space rather than replacing it. Heat-bath configuration interaction (HCI) reaches better than 1~mHa for the all-electron chromium dimer, whose full variational space contains $2\times10^{22}$ determinants\cite{holmes2016heat}; its semistochastic variant (SHCI) attains comparable accuracy for {F$_2$} in 108 orbitals and {Cr$_2$} in 190 orbitals\cite{sharma2017semistochastic}, and related iterative schemes approach the FCI limit by the same selection principle\cite{liu2016ici}. The density matrix renormalization group (DMRG) achieves similar accuracy by a different route, variationally optimizing the wavefunction within the manifold of matrix product states of fixed bond dimension\cite{White1992dmrg,Chan2002dmrg,chan2011dmrg}. Its accuracy is controlled by the bond dimension and extrapolated in the discarded weight, and is most favourable when the orbitals can be ordered so that entanglement across the resulting chain remains bounded; recent GPU-accelerated implementations extend DMRG to active spaces of the size relevant to transition-metal clusters\cite{Menczer2024DMRGFemoco}. Together, SHCI and DMRG define the classical state of the art against which any emerging computational paradigm, quantum or otherwise, must be benchmarked\cite{Williams2020_BenchmarkManyBody}.

The dissociation of the nitrogen molecule (\ch{N2}) and iron--sulfur (Fe--S) clusters are two such benchmarks, and both are more than numerical exercises. \ch{N2} dissociation is the textbook signature of static correlation in bond breaking of the triple covalent bonding. Fe--S clusters, meanwhile, are the catalytic core of nitrogenase and related metalloenzymes central to biological nitrogen fixation and electron transfer, so their electronic structure bears directly on bioinorganic catalyst design; their dense manifold of near-degenerate spin states, arising from strong exchange coupling across multiple Fe centers, places them among the most demanding multireference problems in chemistry. A method that can be trusted on both \ch{N2} and Fe--S clusters is therefore being tested against a problem of fundamental and applied consequence, not merely against a convenient toy system.

Quantum computers offer, in principle, a native representation of correlated many-electron wavefunctions that sidesteps this classical scaling wall\cite{Preskill2018NISQ, Arute2019Sycamore, Kjaergaard2020superconducting}. In practice, however, current noisy intermediate-scale quantum (NISQ) processors are limited by two compounding obstacles: circuit depths sufficient to represent strong correlation quickly exceed coherence times, and the number of measurements required to estimate energies to chemical accuracy grows steeply with system size\cite{McClean2016, Cerezo2021, Bharti2022}. Neither obstacle is resolved by scaling the quantum processor in isolation. Hybrid quantum--classical algorithms address this by dividing workload according to comparative advantage, using the quantum processor only for what it does uniquely well, preparing and sampling correlated configurations, while offloading optimization and subspace diagonalization to classical hardware\cite{Peruzzo2014, McClean2016, robledo2025quantumcentric}.

\begin{figure}[ht]
    \centering
    \includegraphics[width=0.8\linewidth,]{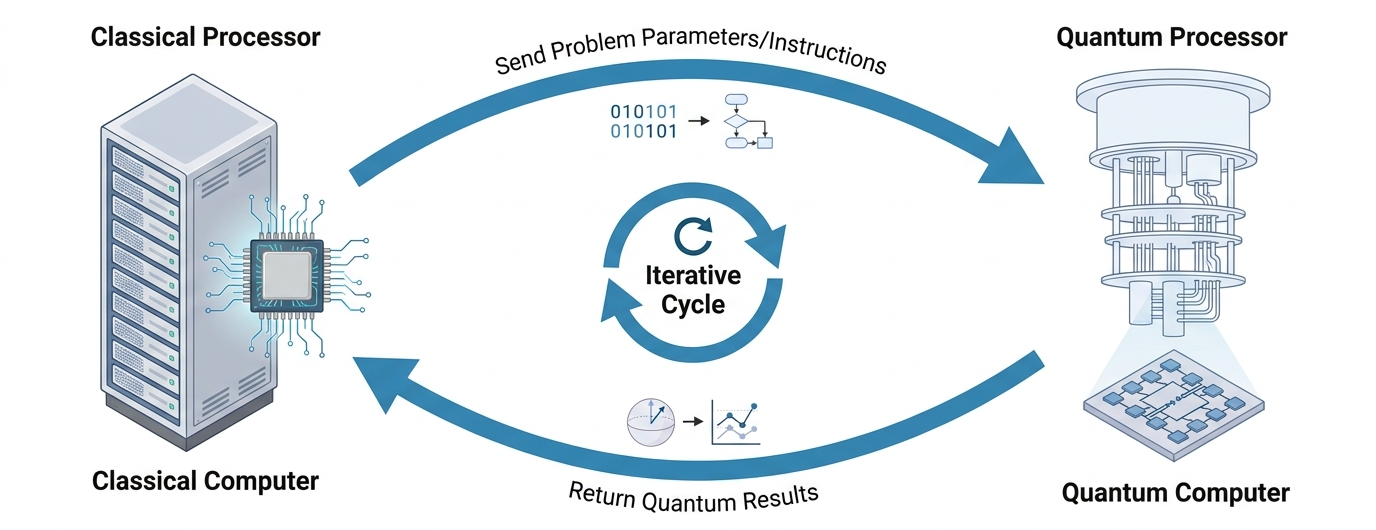}
    \caption{Quantum--classical resource architecture. The quantum processor is used exclusively to propose correlated electronic configurations; all subspace construction, diagonalization, and parameter optimization are executed on the classical partition. The two resources exchange information once per handover iteration.}
    \label{fig:architecture}
\end{figure}

This division of work is also the reason why a quantum-only or classical-only strategy cannot reach the systems studied here. The [2Fe--2S] and [4Fe--4S] clusters require the eigenvalue of the $10^{8}$--$10^{15}$ determinants for the exact solution with a given active space, far beyond the exact or even SCI diagonalization at reasonable cost, while their electronic structure is too entangled for a shallow NISQ-compatible circuit to represent unassisted. Making this hybrid division of work at a scale that actually matters for such systems requires pairing large-scale quantum hardware with comparably large classical resources: in this work, we couple IBM quantum hardware to the supercomputer Fugaku, one of the world's largest quantum--classical computing hybrid systems. Figure~\ref{fig:architecture} summarizes the architecture of the quantum-classical hybrid system.

The variational quantum eigensolver (VQE) has been among the leading candidates for chemistry on noisy intermediate-scale quantum (NISQ) devices\cite{Peruzzo2014,McClean2016,Tilly2022}. A parameterized circuit prepares a correlated trial state on hardware, its energy is estimated from repeated measurements, and a classical optimizer updates the parameters to lower that energy; because only the state preparation is quantum, the coherence-time requirements are far milder than those of quantum phase estimation. Ansatz design has advanced considerably, including adaptive constructions that grow the circuit one operator at a time\cite{Grimsley2019}. In practice, however, VQE applications have remained confined to a few tens of qubits and active spaces of roughly two to sixteen orbitals. Three obstacles compound: the number of measurements required to estimate an energy to chemical accuracy\cite{Izmaylov2020Pauli}, the sensitivity of that estimate to hardware noise, and the flattening of the optimization landscape as the circuit grows\cite{Mcclean2018,Cerezo2021}.

More recently, sample-based quantum diagonalization (SQD) has reorganized this division of labor. The quantum processor is used only to propose candidate electronic configurations: the circuit is sampled in the computational basis, and bitstrings corrupted by noise into the wrong particle-number sector are repaired against the average orbital occupancies in a self-consistent configuration-recovery loop. The Hamiltonian is then diagonalized classically in the subspace spanned by the recovered configurations, yielding a variational upper bound that can be certified classically. Introduced as quantum-selected configuration interaction\cite{kanno2023quantum}, the approach has been carried beyond the scale of exact diagonalization on \ch{N2} and Fe--S clusters using a quantum-centric architecture that couples IBM quantum hardware to the Fugaku supercomputer\cite{robledo2025quantumcentric,shirakawa2025closedloopcalculationselectronicstructure}.

The Handover-Iterative VQE (HI-VQE)\cite{pellowjarman2025hivqe} adopts the same architectural logic but closes the loop around the circuit itself. Whereas SQD draws its configurations from a circuit whose parameters are fixed in advance, HI-VQE alternates between the two solvers: at each iteration the quantum device proposes new Slater determinants, the classical solver diagonalizes the Hamiltonian in the subspace they span, and the resulting eigenvector defines the objective against which the circuit parameters are re-optimized for the following iteration. In parallel, all proposed determinants are retained in an accumulated subspace that supplies the reported variational energy, so the expansion is refined progressively rather than drawn once and fixed (Fig.~\ref{fig:HI-VQE}). The scheme has been benchmarked on \ch{NH3}, \ch{Li2S} and \ch{N2}, reproducing reference CASCI, HCI and DMRG energies; because no energy is estimated from Pauli measurements on hardware, it avoids the measurement overhead and the attendant vanishing-gradient behaviour that limit energy-based VQE\cite{pellowjarman2025hivqe}.

Here, we extend HI-VQE to the scale required to test it against the hardest classical benchmarks available. Coupling IBM quantum hardware to Fugaku, we run HI-VQE on \ch{N2} dissociation (56 qubits) and on the [2Fe--2S] (40 qubits) and [4Fe--4S] (72 qubits) clusters, the latter the largest system reported for this algorithm to date. Benchmarked against HCI and DMRG, HI-VQE reproduces the \ch{N2} potential energy surface to sub-millihartree accuracy while using a determinant subspace one to two orders of magnitude more compact than HCI, indicating that quantum-informed sampling identifies high-weight determinants that a hierarchical excitation tree reaches only at much greater expense. On [4Fe--4S], we report this alongside the residual $\sim\!64$~mHa gap to DMRG, and the significance of the [4Fe--4S] result lies in the fraction of correlation recovered, and we achieve equivalent accuracy with the current state of the art.

Independently of these energy estimation results, we report the full resource usage of one of the calculations on the Fugaku HPC cluster: approximately $0.66$ million node-hours, together with the per-iteration quantum sampling, subspace dimensions, sparse-Hamiltonian sizes, and peak memory footprints. Yet such resource accounting is largely absent from the hybrid-algorithm literature, making it difficult to assess actual costs or to plan future deployments. A concrete accounting of the quantum and classical resources genuinely required to push a hybrid algorithm past exact-diagonalization scale on a chemically meaningful system is therefore a contribution in its own right, of direct use in dimensioning the next generation of quantum-centric supercomputing architectures.

The remainder of this paper is organized as follows.  We first describe the active-space construction and computational setup, then present HI-VQE results for \ch{N2} dissociation and the [2Fe--2S] and [4Fe--4S] clusters benchmarked against HCI, TrimCI, and DMRG.  We conclude with a detailed accounting of quantum and classical resources and an outlook on hybrid algorithms at industrially relevant scale.

\section{Methods}
\subsection{Algorithmic Details of HI-VQE}

\begin{figure}[htbp]
    \centering
    \includegraphics[width=0.6\linewidth,]{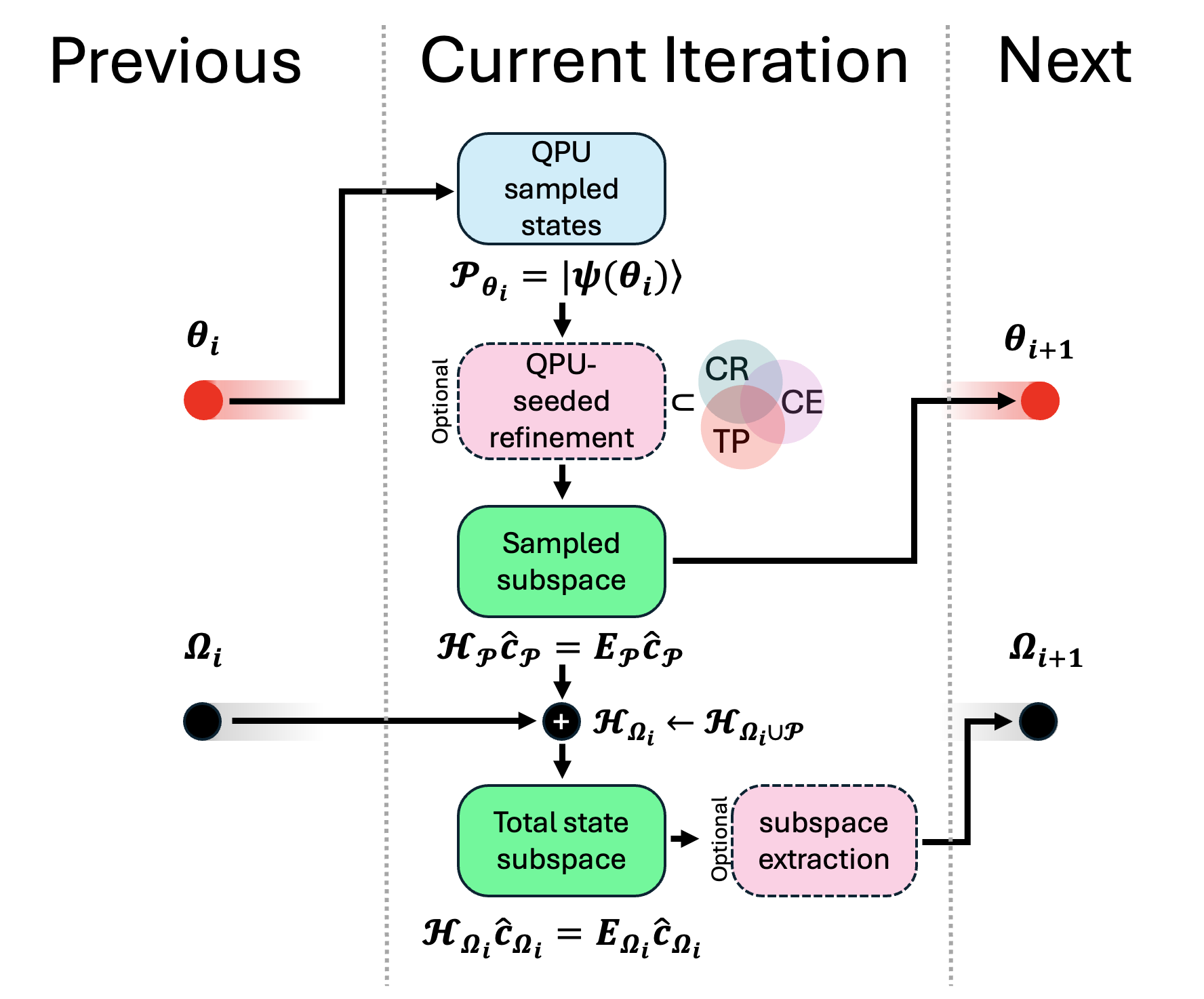}        
    \caption{Overview of the HI-VQE iterative workflow. Each iteration uses QPU-sampled states with the current circuit parameters, optionally refines them via configuration recovery (CR)/tensor product (TP)/classical expansion (CE), and diagonalizes the sampled subspace to update quantum circuit parameter while accumulating subspaces over iterations to obtain the ground state wavefunction. The subspace extraction after diagonalization of accumulated subspace is optionally activated to carry on highly contributing states to obtain highly contributing states rather than keeping all states over the iterations.}
    \label{fig:HI-VQE}
\end{figure}

The basic algorithmic implementation of HI-VQE closely follows the formulation presented in \cite{pellowjarman2025hivqe}, with several refinements incorporated to improve its computational accuracy and efficiency. Figure~\ref{fig:HI-VQE} presents an overview of the modified iterative workflow. In the original method, additional singly and doubly excited configurations generated from the QPU samples were included to recover readily accessible configurations that may have been missed during sampling. As the system size increased, typically exceeding 40 qubits, this simple expansion scheme led to a rapidly growing subspace and became computationally demanding to achieve a desired level of accuracy, thereby limiting the scalability of the method.

Instead of this naive approach, three sequential but optional refinement steps were adopted. (1) \textit{configuration recovery}, proposed in \cite{robledo2025quantumcentric}, was incorporated as an error-mitigation step for noisy QPU samples. This procedure corrects invalid electronic configurations based on the electronic occupation numbers obtained from the diagonalization of the core state in the preceding iteration. For the initial iteration, the Hartree--Fock occupation numbers or those of another designated reference state were used as the baseline for correction. (2) \textit{tensor product} was employed to expand the sampled subspace by separating each electronic configuration into its $\alpha$- and $\beta$-spin components and forming new configurations from their tensor products. For closed-shell systems, the $\alpha$- and $\beta$-spin configurations can be combined into a common set prior to constructing the tensor products, thereby generating additional relevant configurations without further QPU sampling. This procedure can reduce the required quantum sampling effort, although potentially at the cost of increasing the dimension of the resulting subspace. (3) \textit{classical expansion} was finally applied as an HCI-inspired refinement step to further enlarge the sampled subspace. Following the HCI selection strategy \cite{sharma2017semistochastic}, additional determinants generated through single and double excitations were retained when they satisfied the criterion $|H_{ai}c_i| > \epsilon_1$, where $H_{ai}$ denotes the Hamiltonian matrix element coupling a candidate determinant $a$ to a reference determinant $i$, and $c_i$ is the corresponding CI coefficient. This criterion preferentially selects determinants that are strongly coupled to QPU-sampled configurations with large amplitudes, thereby enabling a targeted expansion while maintaining strong Hamiltonian connectivity within the sampled subspace.

Following these sampling-refinement steps, the resulting \textit{sampled subspace} was diagonalized, and an effective subspace was extracted by applying an amplitude-based truncation with a user-defined threshold. A subsequent \textit{total state subspace} diagonalization was then performed in the space spanned by the accumulated $\Omega$ set together with the selected effective subspace. A second amplitude-based truncation was applied to the resulting CI coefficients to identify important configurations, which were added to, rather than used to redefine, the existing $\Omega$ set. Consequently, $\Omega$ grows cumulatively over successive iterations, preserving configurations identified as important in previous iterations. This cumulative update stabilizes the evolution of the variational subspace, reducing abrupt iteration-to-iteration fluctuations and resulting in a smoother, more progressive decrease in the energy throughout the iterative procedure.

\begin{figure}[htbp]
    \centering
    \includegraphics[width=0.75\linewidth,]{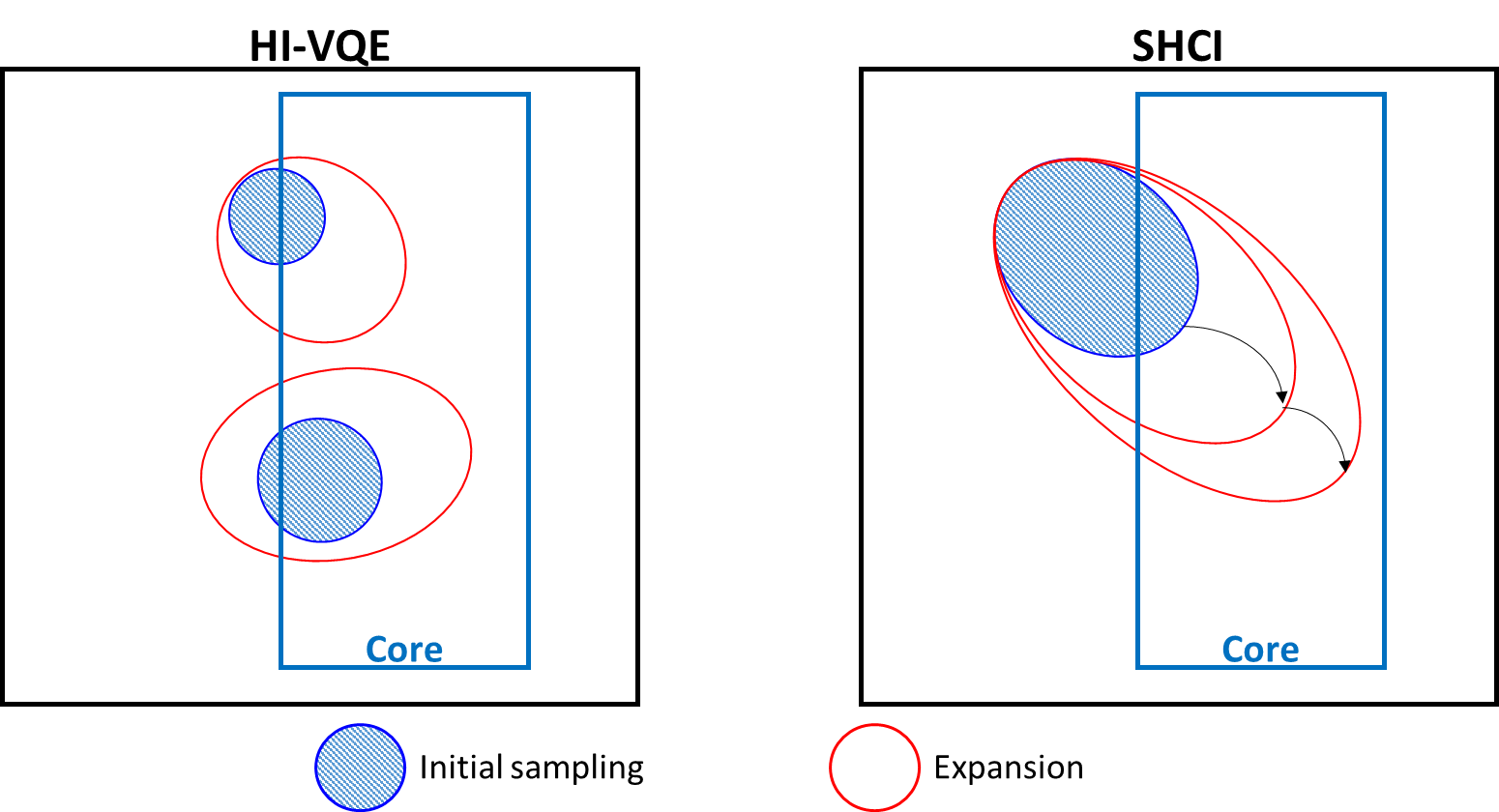}
    \caption{Subspace expansion behavior of left) HI-VQE, and right) HCI}
    \label{fig:Subspace_expansion}
\end{figure}

Although the criterion is defined identically, the subspace expansion mechanisms of HI-VQE and HCI differ fundamentally in how additional configurations are generated and selected, as illustrated in Figure~\ref{fig:Subspace_expansion}. In HCI, the variational space is expanded selectively starting from the current set of important determinants, the expansion therefore proceeds preferentially along strongly connected directions in the Hilbert space, allowing the variational space to grow in a sparse and targeted manner. Newly selected determinants can subsequently serve as additional expansion centers, resulting in a progressive propagation through strongly coupled configurations. While this strategy is highly effective in limiting the size of the variational space, its exploration is inherently guided by the connectivity to configurations that have already been identified as important. In contrast, the HI-VQE expansion is seeded directly by configurations obtained from QPU sampling. Because the quantum state naturally contains contributions from correlated higher-order excitations, QPU sampling can directly access configurations that would otherwise require multiple successive excitation and selection steps in a purely classical expansion scheme. This allows HI-VQE to explore distant but physically relevant regions of the Hilbert space within relatively few iterations. The subsequent local expansion around these sampled configurations further recovers nearby determinants, enabling efficient coverage of a broad portion of the relevant Hilbert space. This can accelerate the discovery of important correlated configurations that are less readily reached through connectivity-driven classical expansion alone.

\subsection{Computational Details}
\begin{table}[htbp]
    \centering
    \caption{Chemical systems and their computational parameters}
    \label{tab:computationalParameter}
    \begin{center}
    \begin{minipage}{0.9\textwidth}
    \begin{subfigure}[b]{0.30\textwidth}
        \centering
        \includegraphics[width=\textwidth]{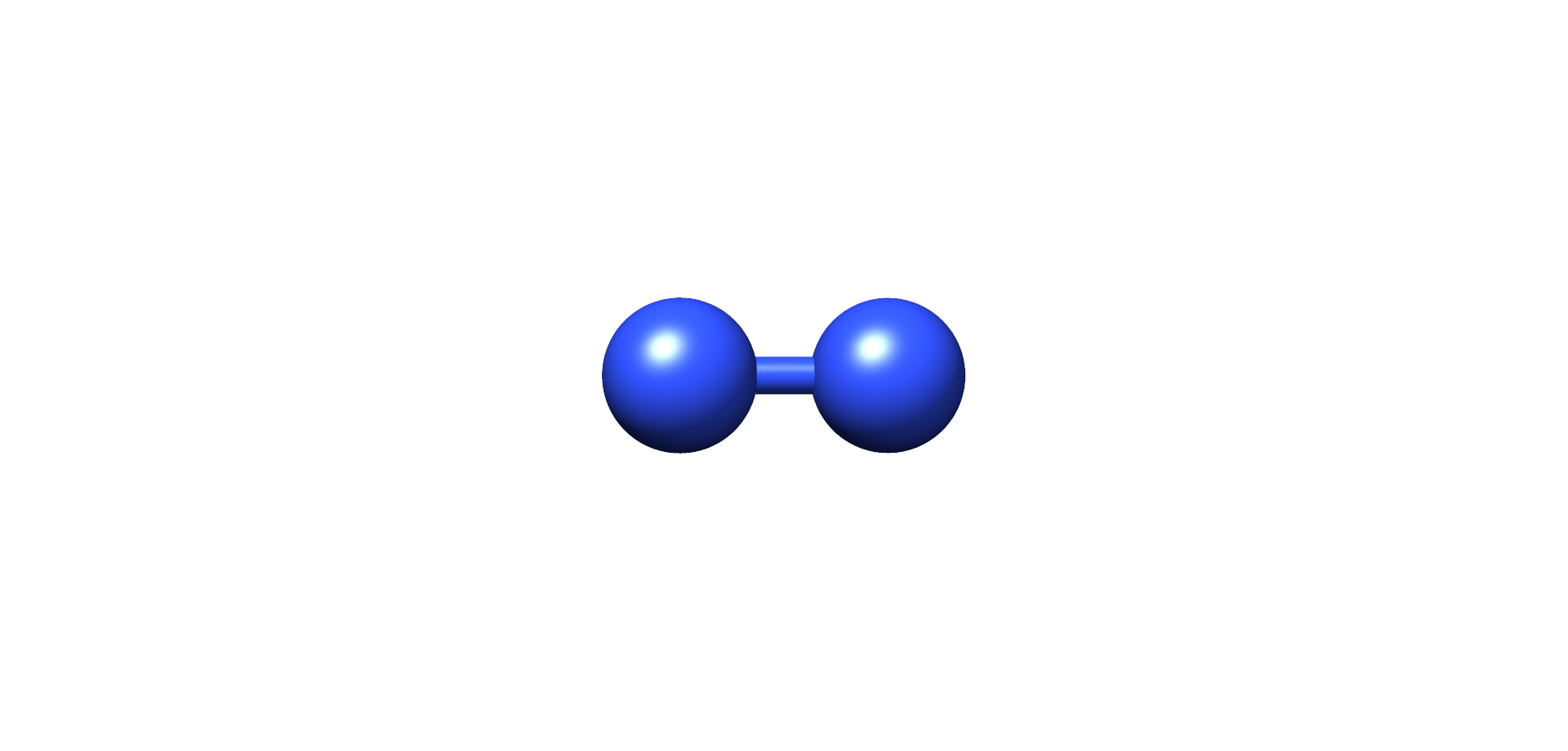}
        \caption{\ch{N2}} 
    \end{subfigure}
    \begin{subfigure}[b]{0.30\textwidth}
        \centering
        \includegraphics[width=\textwidth]{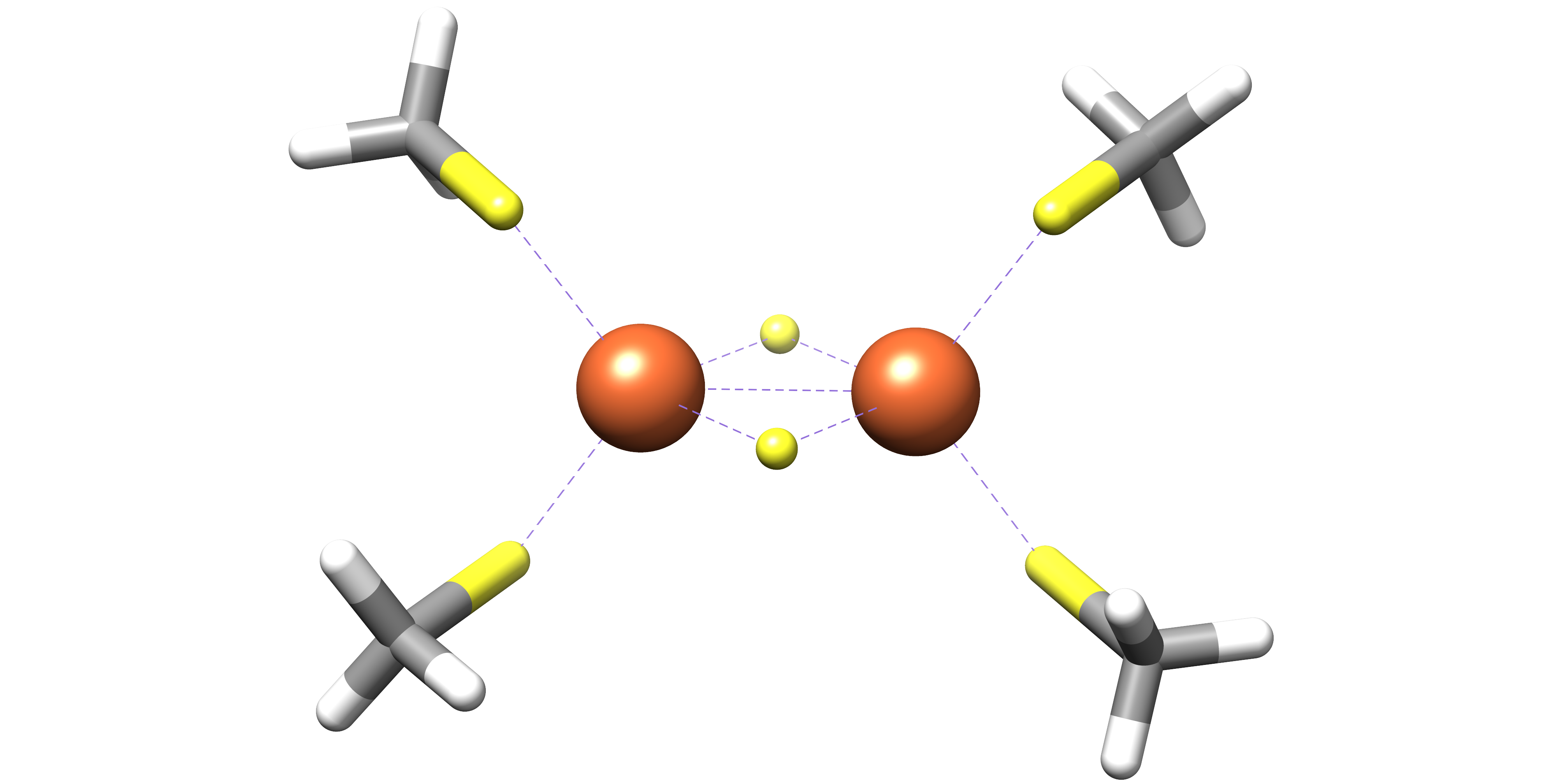} 
        \caption{[2Fe-2S] cluster} 
    \end{subfigure}
    \begin{subfigure}[b]{0.30\textwidth}
        \centering
        \includegraphics[width=\textwidth]{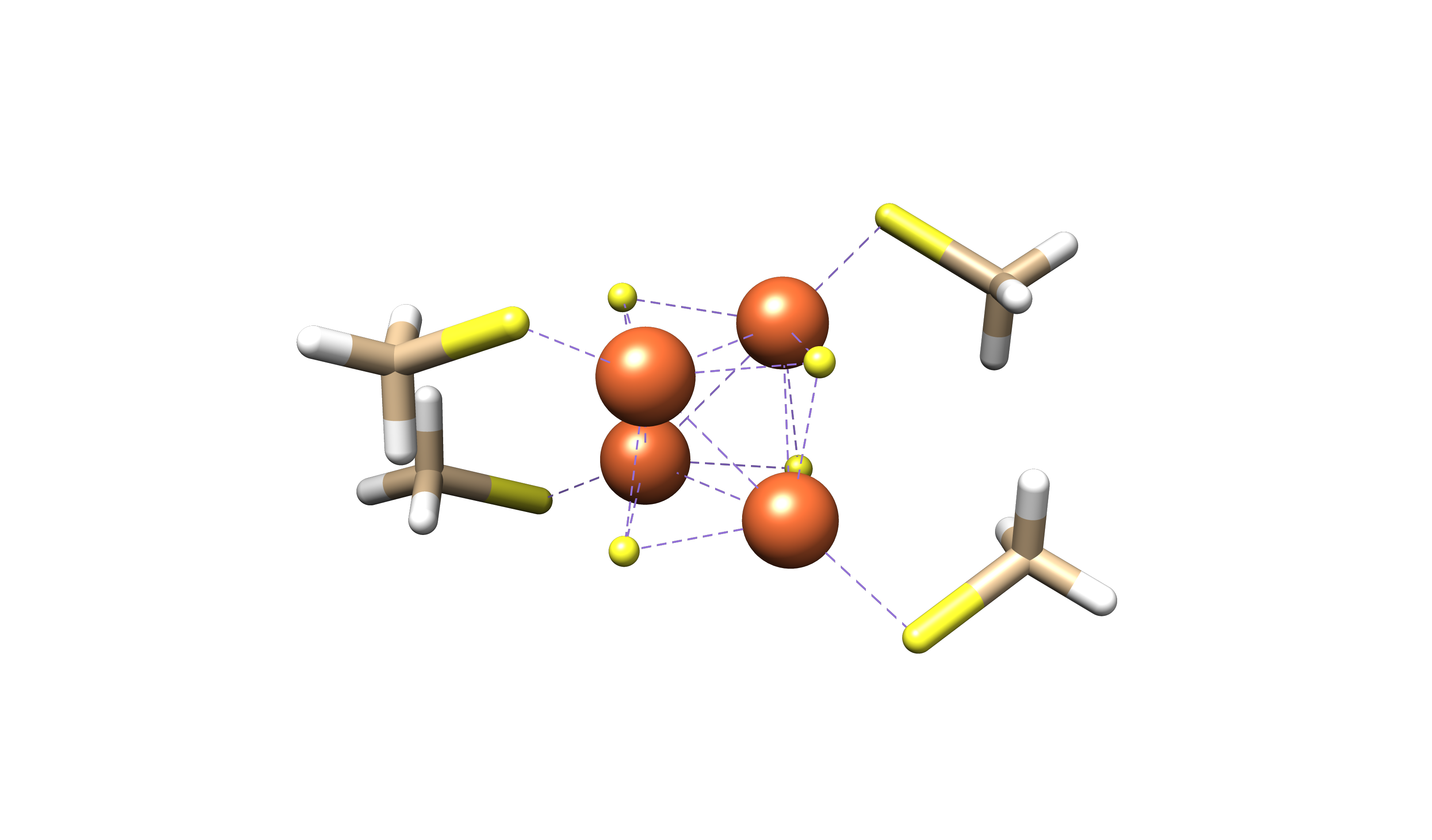} 
        \caption{[4Fe-4S] cluster} 
    \end{subfigure}
    \end{minipage}
    \end{center}
    
    \begin{center}  
    \begin{threeparttable}
    \begin{tabular}{ccccc}
        \toprule
        Chemical Systems & Active Space & CAS size & No. of Qubits & Basis Set \\
        \midrule
        \ch{N2} & (14e,28o)\tnote{a} & $1.40\times10^{12}$ & 56 & cc-pVDZ \\
        \ch{[Fe2S2(SCH3)4]^2-} & (30e,20o) & $2.40\times10^{8}$ & 40 & TZP-DHK \\
        \ch{[Fe4S4(SCH3)4]^2-} & (56e,36o) & $9.15\times10^{14}$ & 72 & TZP-DHK \\
        \bottomrule
    \end{tabular}
    \begin{tablenotes}
        \item[a] all-electron calculation
    \end{tablenotes}
    \end{threeparttable}
    \end{center}
\end{table}

The molecules simulated in this work are summarized in Table \ref{tab:computationalParameter}. The dissociation of \ch{N2} is studied using the cc-pVDZ basis and an all-electron active space with (14e, 28o). As a classical reference, the potential energy curve was computed using the HCI method. The molecular geometry and active space selection of the \ch{[Fe2S2(SCH3)4]^2-} and \ch{[Fe4S4(SCH3)4]^2-} clusters (denoted as [2Fe–2S] and [4Fe–4S]) used in this work were adopted from Robledo-Moreno et al. 2025 \cite{robledo2025quantumcentric}. This ensures that our computational results remain directly comparable to the benchmarking data reported in that study, facilitating a rigorous assessment of the accuracy and convergence of the method. Specifically, the atomic coordinates of the [2Fe–2S] and [4Fe–4S] clusters were taken as reported in \cite{sharma2014natchem} and the active space was selected using the procedure described in \cite{li2017spinprojected} and available on the SQD data repository \cite{SQD_data_repo}.

\begin{table}[htbp]
    \centering
    \caption{Molecular structures and Parameters for HI-VQE calculations} %
    \label{tab:ComputationalConditions}
    \begin{tabular}{cccccc}
        \toprule
        Calculation type & Backend & No. of Shot & No. of Qubit & Ansatz & No. of Calculations \\
        \midrule
        \ch{N2:(Dist,Energy)} & \texttt{ibm\_fez} & 4,000 & 56 & EPA-L-R2 & 9 \\
        \ch{N2:(Qubits,Dets)} & \texttt{ibm\_fez} & 4,000 & 24--56 & EPA-L-R2 & 9 \\
        \text{[2Fe-2S]}:case\_1 & \texttt{ibm\_fez} & 50,000 & 40 & EPA-L-R2 & 3 \\
        \text{[2Fe-2S]}:case\_2 & \texttt{ibm\_fez} & 50,000 & 40 & \textit{Ad hoc} & 3 \\
        \text{[2Fe-2S]}:case\_3 & \texttt{ibm\_fez} & 100,000 & 40 & \textit{Ad hoc} & 1 \\
        \text{[2Fe-2S]}:case\_4 & \texttt{ibm\_fez} & 25,000 & 40 & \textit{Ad hoc} & 1 \\
        \bottomrule
    \end{tabular}
\end{table}

This section summarizes the key computational settings used for all HI-VQE simulations carried out in this work. All quantum-sampling procedures were executed on the \texttt{ibm\_fez} backend of the IBM Quantum platform, covering molecular systems ranging from 24 to 56 qubits. Table~\ref{tab:ComputationalConditions} provides a detailed overview of the configuration employed for each benchmark, including the backend, number of measurement shots, qubit counts, ansatz types, and the number of independent HI-VQE calculations performed.

The \ch{N2} cc-pVDZ benchmarks consist of two categories: (i) 56-qubit Potential Energy Surface (PES) calculations used to investigate dissociation behavior (\ch{N2}:(Dist,Energy)), and (ii) variable-qubit studies (24--56 qubits) designed to examine determinant number growth and resource scaling denoted as \ch{N2}:(Qubits,Dets). 
For the [2Fe–2S] cluster, we evaluated the effect of measurement noise using 50,000-shot ([2Fe-2S]:case\_1) and 100,000-shot ([2Fe-2S]:case\_3) simulations. In addition, the 50,000-shot case was used to compare two quantum Ansätze for feasibility assessment: the EPA-L-R2 excitation-preserving Ansatz ([2Fe-2S]:case\_1) and a tailored \textit{ad hoc} circuit ([2Fe-2S]:case\_2).
The EPA-L-R2 denotes an excitation-preserving Ansatz with linear entanglement and two repetitions of the circuit block.

\section{Results and Discussion} %

\subsection{[2Fe-2S] cluster} \label{[2Fe-2S]_cluster}
\begin{table}[h!]
    \centering
    \begin{threeparttable}[b]
    \captionsetup{width=1\textwidth} 
    \caption{[2Fe-2S] cluster in TZP-DHK basis: Ground state energy, energy error and number of determinants for Hartree-Fock, CCSD, HCI and HI-VQE compared to CASCI reference.} %
    \label{tab:fes_properties}
    \begin{tabular}{cccr}
        \toprule
        Method & E$_{total}$ (Ha) & $\Delta E$\tnote{a} (mHa) & No. of Dets \\
        \midrule
        Hartree-Fock & -116.205816 & 399.7930 & 1 \\
        CCSD\tnote{b} & -116.405662 & 199.7473 & -\\
        HCI(30e,20o) \tnote{b} & -116.604878 & 0.7310 & 56,665,658 \\
        HCI(30e,20o) \tnote{c} & -116.605333 & 0.2760 & 97,503,808 \\
        HCI(30e,20o) \tnote{d} & -116.605409 & 0.2000 & 196,112,016 \\
        CASCI(30e,20o)  & -116.605609 & 0.0000 &  240,374,016  \\
        HI-VQE (30e,20o) & & & \\
        case\_1.1 & -116.604576 & 1.0332 & \num{66861907} \\
        case\_1.2 & -116.604682 & 0.9269 & \num{65456175} \\
        case\_1.3 & -116.604779 & 0.8298 & \num{63857547} \\
        case\_2.1 & -116.604742 & 0.8666 & \num{71377177} \\
        case\_2.2 & -116.604536 & 1.0727 & \num{64449164} \\
        case\_2.3 & -116.604543 & 1.0660 & \num{66330225} \\
        case\_3 &   -116.601723 & 3.8859 & \num{53871015} \\
        case\_4 &   -116.604694 & 0.9151 & \num{72558573} \\
        \bottomrule
    \end{tabular}
    \begin{tablenotes}
        \item[a] $E_m - E_{CASCI}$
        \item[b] Arrow-HCI with $\epsilon=5\times10^{-6}$
        \item[c] Arrow-HCI with $\epsilon=2\times10^{-6}$
        \item[d] Ref. \cite{robledo2025quantumcentric}
    \end{tablenotes}
    \end{threeparttable}
\end{table}

In contrast to  \ch{N2}, Fe–S clusters such as [2Fe–2S], [4Fe–4S], and the FeMo cofactor (FeMoco) in nitrogenase enzymes pose a far more complex problem. 
These clusters, composed of multiple high-spin Fe centers bridged by sulfides, are central to numerous biological processes, including electron transfer, enzymatic catalysis, and nitrogen fixation \cite{beinert1997science,sharma2014natchem,zhang2022oxidmed,zhai2023nitrogenase}. Beyond their biological relevance, Fe–S clusters also underpin emerging applications in biomimetic catalysis, materials design, and quantum sensing further amplifying the need for scalable, accurate electronic-structure solutions. Fe–S clusters serve, therefore, as realistic and chemically meaningful benchmarks for testing next-generation correlated-electron methodologies \cite{hehn2025fes,Zhou2024FeS}. Their dense manifolds of low-lying spin states arise from strong exchange coupling and orbital delocalization, often requiring active spaces of 60–100 orbitals or more to capture essential physics \cite{zhai2023nitrogenase,benediktsson2022fes}. Conventional single-reference approaches such as DFT and coupled-cluster theory (e.g., CCSD(T)) often fail to capture the multireference character intrinsic to Fe–S systems, particularly in low-spin or mixed-valence configurations \cite{Vysotskiy2020_Nitrogenase_Model,Cao2019_Nitrogenase_DFT, Vysotskiy2024_MinimalNitrogenase_CCSDT_errors}. 
To achieve quantitative accuracy, it is necessary to employ multireference wavefunction methods that can systematically recover both static and dynamic correlation within a large active space.  Among classical methods, SHCI, CASCI, and DMRG are widely recognized for their capacity to address the multireference character within large active spaces required for Fe–S clusters \cite{robledo2025quantumcentric, Zhou2024FeS, benediktsson2022fes, sharma2014natchem}. In Fe–S clusters, SHCI has demonstrated remarkable effectiveness in capturing ground and excited state correlation energies together with the self-consistency field (SCF) approach for orbitals optimization \cite{yao2021orbital}. Recent studies show that the energy differences between SHCI and DMRG on Fe–S clusters often approach 1 millihartree, suggesting that both methods can achieve near-exact results when convergence is carefully controlled \cite{sharma2017semistochastic,Abraham2021}.

\begin{figure}[ht!] 
    \centering
    \begin{subfigure}{1.0\textwidth}
        \centering
        \includegraphics[width=0.75\linewidth,]{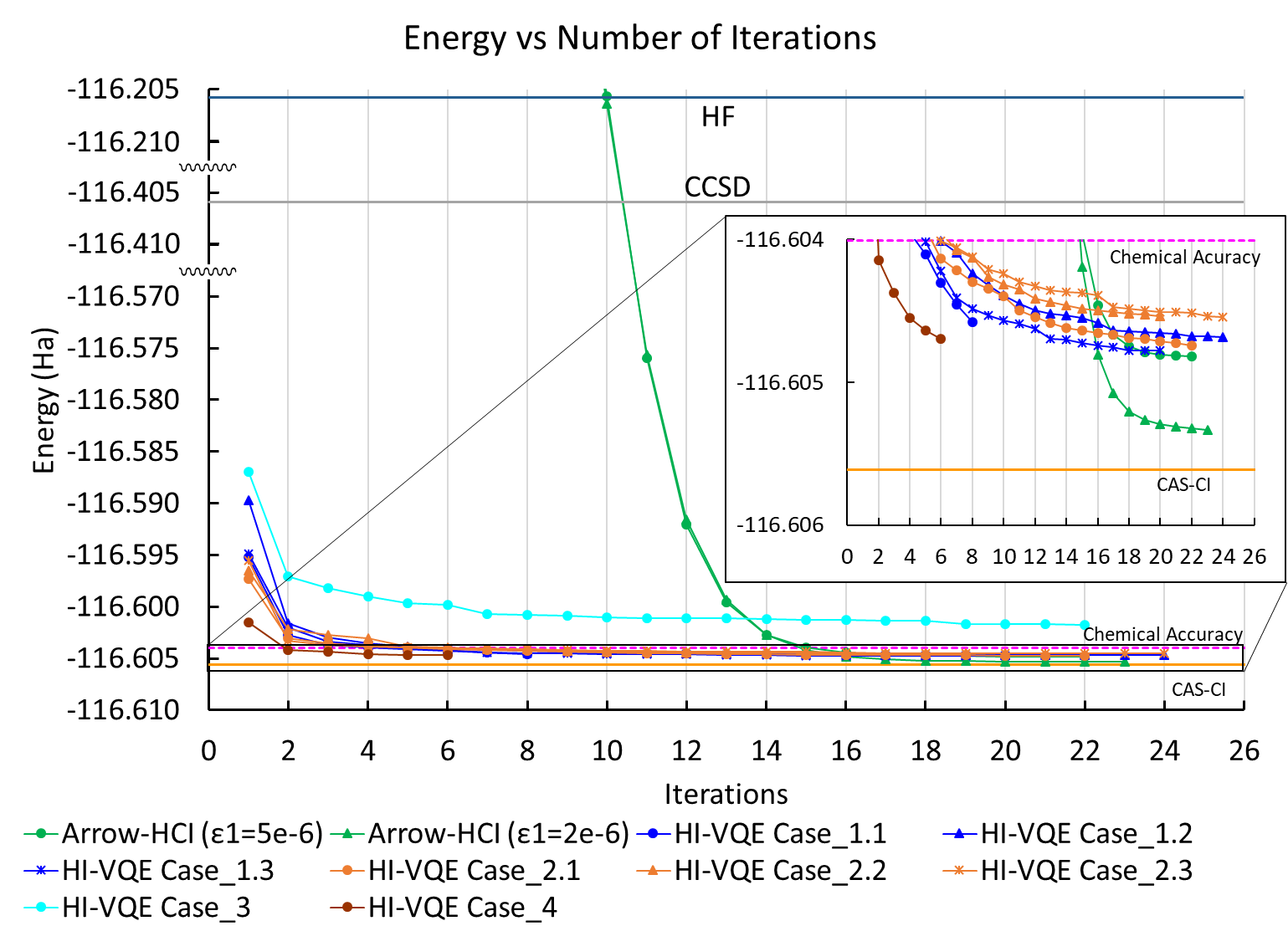} 
        \caption{}
        \label{fig:FeSiterationE}
    \end{subfigure}
    \begin{subfigure}{1.0\textwidth}
        \centering
        \includegraphics[width=0.75\linewidth,]{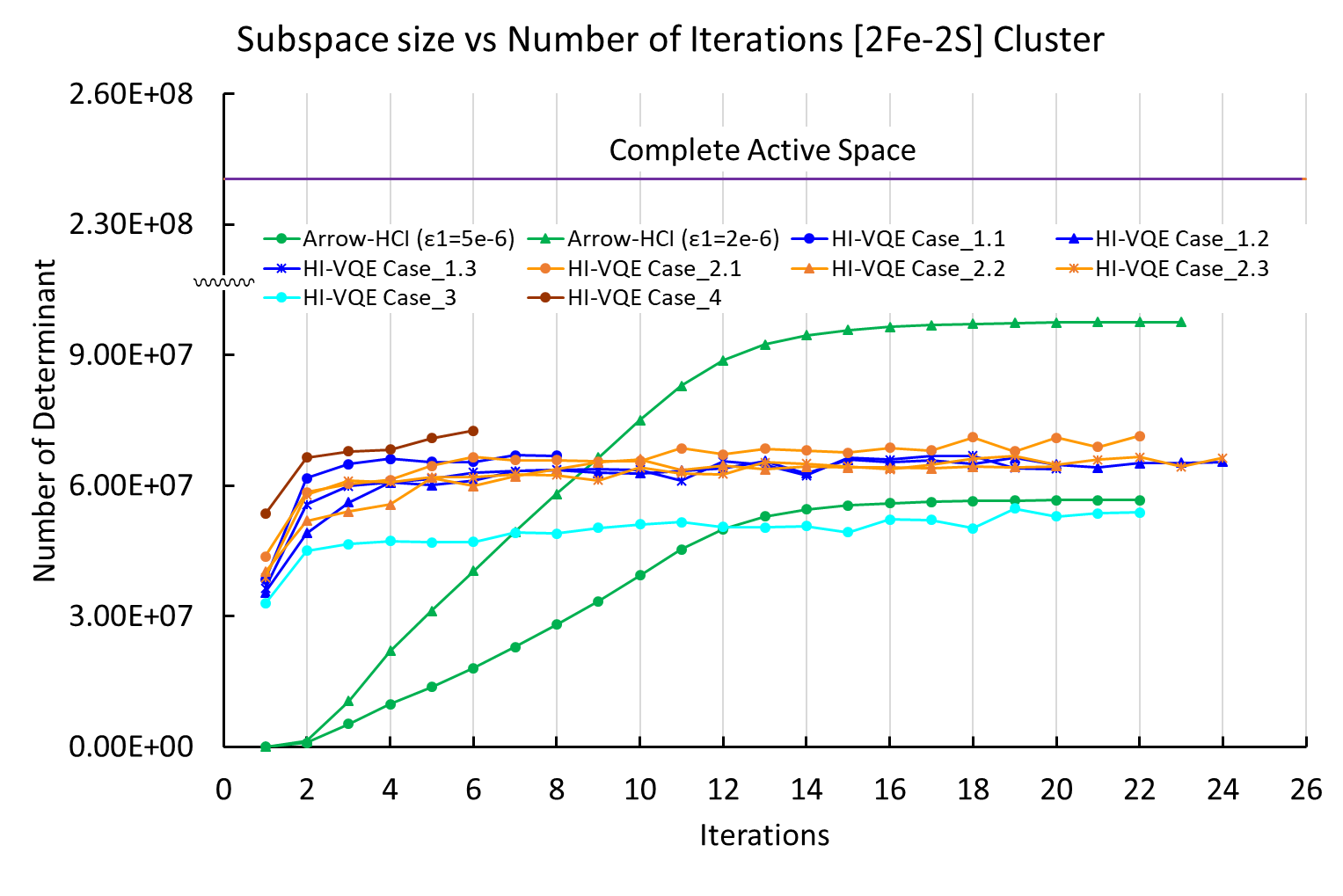} 
        \caption{}
        \label{fig:FeSiteration_subspace}
    \end{subfigure}
    \caption{ (a) The energy vs the number of iterations of [2Fe-2S] cluster (30e,20o) system using the TZP-DHK basis with 40 qubits. The HCI energy was taken from Ref. \cite{robledo2025quantumcentric} and HI-VQE calculations. (b) The subspace size vs the number of iterations of each calculation}
    \label{fig:fes_convergence}
\end{figure}

From a theoretical standpoint, Fe–S clusters represent an extreme case of multireference spin frustration \cite{hehn2025fes,Zhou2024FeS,li2017spinprojected}. Each Fe ion contributes a substantial local magnetic moment, yet the total cluster spin is often low due to antiferromagnetic coupling between centers. 
This interplay of localized and delocalized correlation renders classical FCI or multiconfigurational methods computationally prohibitive. For example, even the [2Fe–2S] cluster, a seemingly simple model, exhibits a singlet ground state emerging from intricate superexchange interactions among spin-coupled and charge-delocalized configurations. The electronic ground state of the [2Fe-2S] cluster consists of two Fe(III) centers with high-spin that are antiferromagnetically coupled. Capturing such phenomena with quantitative accuracy lies beyond the reach of conventional approaches, motivating the exploration of hybrid and quantum-assisted frameworks.

The benchmarking of the Fe–S cluster using HI-VQE reveals a compelling balance between accuracy and computational efficiency. Table \ref{tab:fes_properties} highlights the performance of HI-VQE relative to classical methods such as Hartree–Fock, CCSD, and HCI. While Hartree–Fock and CCSD exhibit large deviations from the CASCI reference, underscoring their inability to capture strong correlation, HCI achieves near-exact results, but at the cost of tens to hundreds of millions of determinants. In contrast, HI-VQE reproduces HCI-level accuracy (0.83 - 3.89 mHa) with roughly one-third of the determinants ($\sim$50–70 million determinants). This reduction in determinant count demonstrates the efficiency of quantum-informed sampling in constructing a compact yet physically meaningful subspace, positioning HI-VQE as a scalable alternative for large and multireference systems.

Figure \ref{fig:FeSiterationE} provides further insight into the iterative behavior of HI-VQE compared to HCI. The energy convergence plot shows that HI-VQE approaches the CAS-CI benchmark within a few iterations, demonstrating the efficacy of its quantum-guided subspace construction. Additionally, the subspace size remains significantly smaller for HI-VQE across all iterations (Figure \ref{fig:FeSiteration_subspace}), even as the algorithm refines the wavefunction. This compactness directly translates into reduced diagonalization costs and improved scalability, which is critical for systems with large active spaces such as Fe–S clusters.

\begin{figure}[ht]
    \centering
    \includegraphics[width=0.75\linewidth,]{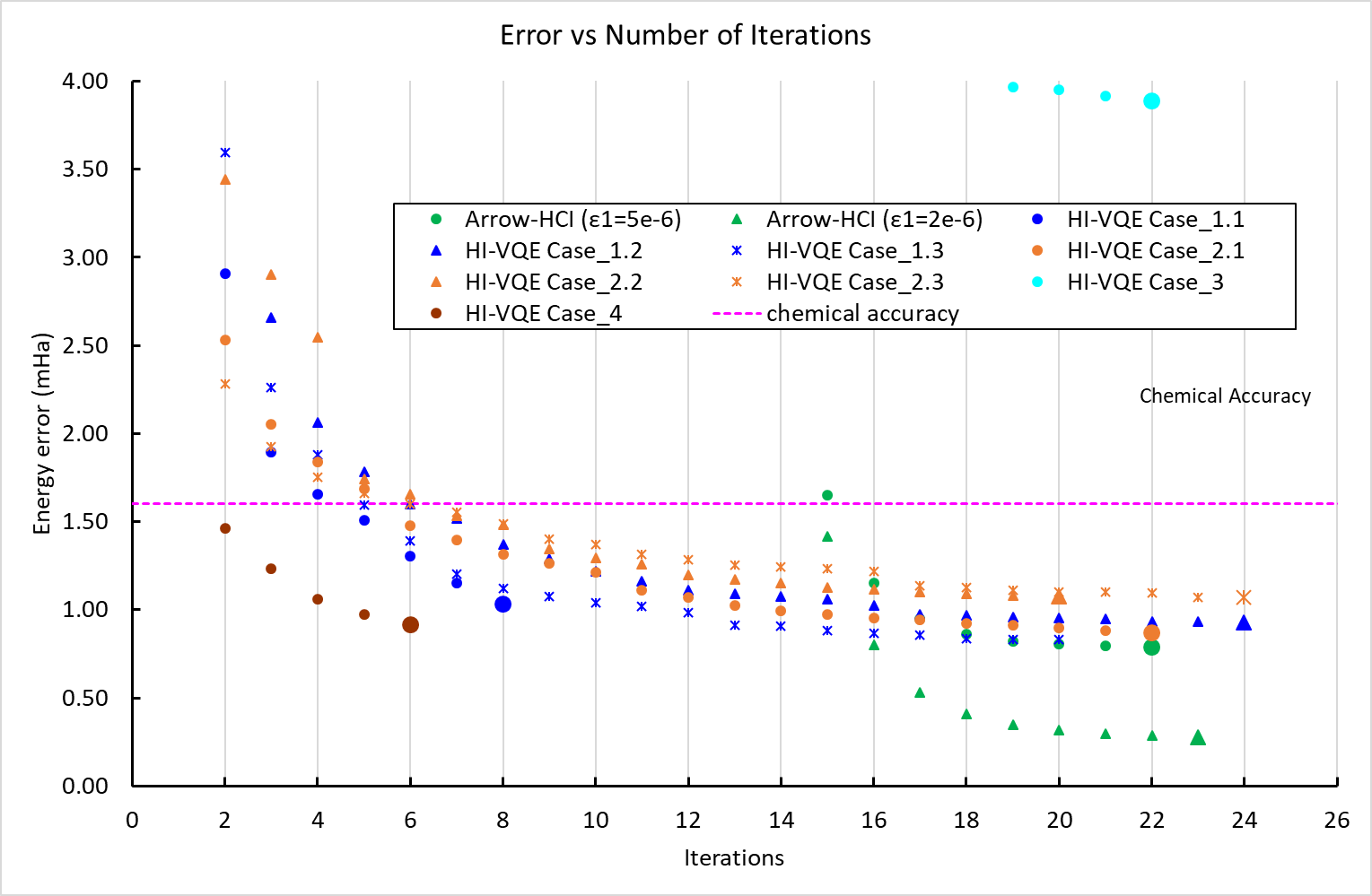} 
    \caption{The energy error vs the number of iterations of [2Fe-2S] cluster (30e,20o) system using the TZP-DHK basis with 40 qubits with respect to the HCI energy taken from Ref. \cite{robledo2025quantumcentric}. }
    \label{fig:fes_energyError}
\end{figure}

Finally, Figure \ref{fig:fes_energyError} tracks the energy error of HI-VQE relative to CAS-CI over successive iterations. The error decreases steadily and approaches chemical accuracy, especially in Case 4, where the number of shots is increased to 100,000. By iteratively optimizing the quantum state from which the samples are obtained, HI-VQE is capable of highly accurate and efficient simulations. These results underscore HI-VQE's potential as a practical quantum–classical hybrid framework for tackling strongly correlated systems, paving the way for applications in bioinorganic chemistry, catalysis, and materials science.

The results reported in Table 5 also demonstrate that the performance of HI-VQE is strongly related to the suitability of the prepared quantum state and sampling procedure. Case 1 (EPA‑L‑R2, 50,000 shots) reaches 0.83 mHa with $\sim$63M determinants, already below the chemical accuracy threshold and orders of magnitude more accurate than single‑reference baselines. The \textit{ad hoc} circuit in Case 2 achieves the same level of accuracy as Case 1 with $\sim$71M determinants, indicating that physically motivated, excitation‑preserving ansätze (EPA‑L‑R2) better exploit the structure of the Fe–S electronic manifold than non‑number‑conserving or less tailored circuits. When the 25,000 shots were applied (Case 3), HI-VQE could not reach the chemical accuracy within 22 iterations. Crucially, Case 4 reduces the calculations to 6 iterations and achieves an error of 0.92 mHa with $\sim$72M determinants by increasing the number of shots to 100,000, underscoring the role of sampling statistics in suppressing stochastic variance and enhancing determinant selection. Across these runs, the determinant savings relative to HCI, which requires 56.7–196.1M determinants to reach $\leq$ 1 mHa, are consistent and substantive, highlighting a clear computational advantage.

We further compare HI-VQE with IBM's sample-based quantum diagonalization (SQD) \cite{robledo2025quantumcentric}, a closely related quantum-classical approach. In particular, for this system HI-VQE reaches 0.83 mHa with $\sim$63M determinants, compared with 151.9 mHa and $\sim$56M determinants for SQD \cite{robledo2025quantumcentric}. This comparison highlights the distinction between the two approaches: HI-VQE relies on an iterative refinement of the subspace through repeatedly updating and sampling the quantum state, whereas SQD samples a single quantum state and relies on iterative classical post-processing routines. The improved accuracy/subspace trade-off observed for HI-VQE demonstrates the benefit of iterative quantum-guided selection for strongly correlated systems.

Overall, the [2Fe--2S] benchmark demonstrates that HI-VQE can retain near-CASCI accuracy while operating with a substantially reduced determinant space compared to HCI. Together with the comparison to SQD, these results indicate that quantum-guided subspace construction can provide a practical route to reducing the computational cost of large active-space calculations.

\subsection{Dissociation of Nitrogen molecule (\ch{N2})} %
The \ch{N2} dissociation curve in the cc-pVDZ basis remains a textbook test for multireference accuracy, as the system transitions from a single-reference description near equilibrium to a highly degenerate manifold at large bond separations.\cite{MALMQVIST1989189,Booth2011JChemPhys} Capturing this transition continuously and quantitatively remains a benchmark of any correlated-electron approach, classical or quantum. Near the equilibrium bond distance, the CCSD and CCSD(T) methods yield potential energies close to experimental dissociation energies due to a well-described dynamic correlation. However, as the N-N bond elongates toward dissociation, the single-reference wavefunction underlying CCSD and CCSD(T) becomes inadequate since it can not describe separated fragments with localized spins, leading to a systematic overestimation of the energy and a failure to capture static correlation effects from near-degenerate electronic configurations. \cite{RishiJCP2016_N2_ccsd,KatsJCP2014_N2_ccsd, Kowalski2000_CCSDT}. Standard density-functional approximations also suffer from the inability to describe systems with fractional spins and fail to reproduce molecular dissociation curves (including \ch{N2}) due to strong static correlation and delocalization errors.\cite{Cohen2008_StaticCorrDFT,Boyn2022_N2DissociationDFTfailure} This highlights a central lesson in quantum chemistry that method choice must track the evolving electron correlation character along the entire PES. In Table \ref{tab:n2_properties}, the experimental dissociation energy of \ch{N2} is compared to the numerical estimations obtained with different computational methods. The dissociation energy was estimated from $E_{diss}=E_{extended} - E_{eq}$ for the computational approaches.

\begin{table}[ht]
    \centering
    \begin{threeparttable}[b]
    \caption{Comparison of HI-VQE and HCI Energies along the N-N distances} %
    \label{tab:benchmarking_energies}
    \begin{tabular}{ccccc}
        \toprule
        \multirow{2}{*}{N-N distances (\AA)} &\multirow{2}{*}{ HI-VQE Energy (Ha)} & \multicolumn{2}{c}{HCI Energy (Ha)}& \multirow{2}{*}{$\Delta$E\tnote{a} (mHa)} \\
 & &  $\epsilon = 2\times 10^{-6}$ &$\epsilon = 1\times 10^{-6}$&\\
        \midrule
        0.90 & -109.0625337 & -109.062579  & -109.062600 & 0.0667 \\ 
        0.80 & -108.6659061 & -108.665944  & -108.665961 & 0.0552 \\ 
        1.00 & -109.2322559 & -109.232352  & -109.232376 & 0.1197 \\ 
        1.10 & -109.2810451 & -109.281095  & -109.281122 & 0.0771 \\ 
        1.20 & -109.2684836 & -109.268547  & -109.268575 & 0.0910 \\ 
        1.50 & -109.1284202 & -109.128515  & -109.128547 & 0.1268 \\ 
        2.00 & -108.9849804 & -108.985157  & -108.985189 & 0.2083 \\ 
        2.50 & -108.9640293 & -108.964173  & -108.964199 & 0.1698 \\ 
        3.00 & -108.9612287 & -108.961473  & -108.961496 & 0.2675 \\ 
     
        \bottomrule
    \end{tabular}
    \begin{tablenotes}
        \item[a] $E_{HI-VQE}$ - $E_{HCI(\epsilon = 1\times 10^{-6})}$ 
    \end{tablenotes}
    \end{threeparttable}
\end{table}

\begin{figure}[ht!]
    \centering
    \includegraphics[width=0.6\linewidth,]{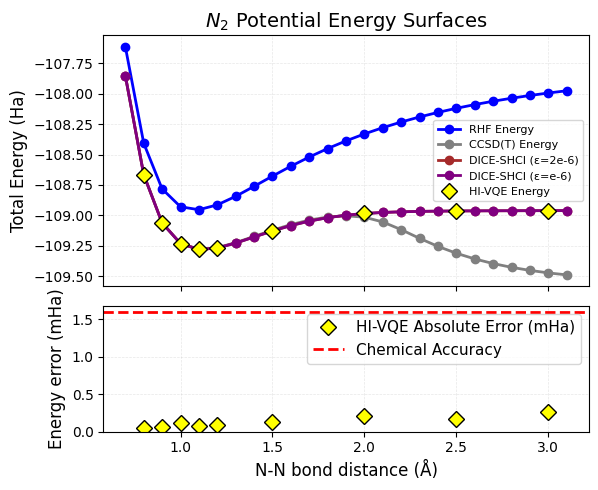}        
    \caption{Top: Potential energy surface of \ch{N2} cc-pVDZ calculated by RHF, HCI and HI-VQE. Bottom: HI-VQE energy error w.r.t. HCI $(\epsilon = 1\times 10^{-6})$}
    \label{fig:N2_PES}
\end{figure}

The performance of the HI-VQE for the \ch{N2} potential energy surface was systematically investigated by benchmarking against HCI energies throughout the dissociation coordinates. Figure \ref{fig:N2_PES} (top panel) presents the potential energy surfaces (PES) for \ch{N2} as calculated by both HCI (purple) and HI-VQE (yellow points) over a range of N–N bond distances (0.8 - 3.0 \AA). The data demonstrate excellent agreement between the two methods, and HI-VQE successfully captures both the equilibrium and the dissociation regime of the PES. The lower panel of Figure \ref{fig:N2_PES} quantifies the absolute errors of the HI-VQE energies relative to HCI at each bond length, expressed in millihartree (mHa). The HI-VQE method achieves agrements with HCI well below the millihartree, as shown in Table \ref{tab:benchmarking_energies}. This is a key benchmark, indicating that HI-VQE not only reproduces quantitative energetics at equilibrium but also maintains high fidelity across the full dissociation curve, an essential capability for accurately modeling bond breaking and strong electron correlation.

\begin{figure}[ht!]
    \centering
    \includegraphics[width=0.6\linewidth,]{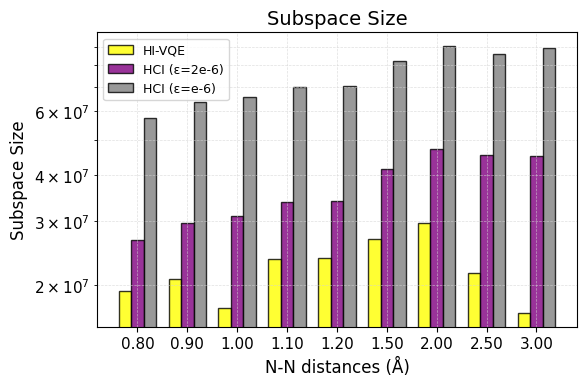} 
    \caption{\ch{N2} cc-pVDZ dissociation: subspace size comparison between HCI and HI-VQE.}
    \label{fig:N2_subspace}
\end{figure}

In the strongly correlated stretched regime, the multireference character of the wavefunction requires a larger number of determinants for an accurate representation. These determinants must be carefully selected to capture the correlation effects. In Figure \ref{fig:N2_subspace}, the subspace size (i.e., the number of determinants or electronic configurations) is shown as a function of the bond length. HI-VQE consistently operates with a significantly reduced subspace size compared to HCI at all distances. At stretched bond lengths (e.g., 3.00 \AA), the computational burden for HCI escalates sharply, requiring subspaces approaching $4.5\times10^7$, whereas HI-VQE maintains a substantially lower resource demand.
These results indicate that HI-VQE performs a more physically meaningful and compact selection of contributing determinants in the strongly correlated regime compared to HCI.

\begin{table}[ht]
    \centering
    \begin{threeparttable}[b]
    \captionsetup{width=1\textwidth} 
    \caption{\ch{N2} dissociation energy from experiments and different computational approaches} %
    \label{tab:n2_properties}
    \begin{tabular}{cc}
        \toprule
        Method & Dissociation energy (eV) \\
        \midrule
        Experiment\tnote{a} & 9.75(6) \\
        HF cc-VDZ\tnote{b} & 26.12 \\
        DFT B3LYP cc-pVQZ\tnote{c,d}& $>$ 16 \\
        CCSD(T) cc-pVDZ \tnote{b,d} &  -- \\
        HCI cc-pVDZ\tnote{d} & 8.70 \\
        HI-VQE cc-pVDZ\tnote{d} & 8.70 \\
        \bottomrule
    \end{tabular}
    \begin{tablenotes}
        \item[a] Ref. \cite{FrostDC1956N2diss,PengWJchemphys2024N2diss}
        \item[b] Ref. \cite{Cohen2008_StaticCorrDFT}
        \item[c] Ref. \cite{RishiJCP2016_N2_ccsd}
        \item[d] Estimated by $E_{\mathrm{diss}}$ = $E_{\mathrm{Extended}} - E_{\mathrm{eq}}$
    \end{tablenotes}
    \end{threeparttable}
\end{table}

\subsection{Efficient Wavefunction Approximation using HI-VQE}
SCI methods, including HCI, rely on a hierarchical tree search in which configurations are generated by applying increasing excitation ranks to a single reference determinant, typically Hartree–Fock. Although SCI achieves better practical scaling than FCI, often exhibiting sub-exponential growth of the selected subspace, the search process is fundamentally constrained by the sparse nature imposed by the Slater–Condon rules. Each determinant is meaningly connected only to a limited set of other determinants through single and double excitations, creating a sparse and shallow connectivity graph. As a result, important determinants that lie far from the HF reference in excitation space, and which are crucial to describe strongly correlated multireference regimes \cite{Ball2022_StrongCorrelationHighExcitations}, are difficult to reach through this restricted excitation tree. The method must therefore evaluate many candidate excitations at each level. The number of accessible configurations grows combinatorially with system size. This makes it increasingly difficult to efficiently identify a compact yet high-weight determinant set. By contrast, HI-VQE leverages QPU samples to directly probe the structure of the correlated wavefunction. The QPU generates excited electronic configurations without being restricted to excitation rank or proximity to the reference states, enabling expanded determinant sets to be seeded directly from measured bitstrings. As a result, HI-VQE naturally uncovers a more compact and physically relevant determinant subset by avoiding the tree-search bottleneck intrinsic to classical SCI methods.

\begin{figure}[ht]
    \centering
    \includegraphics[width=0.6\linewidth,]{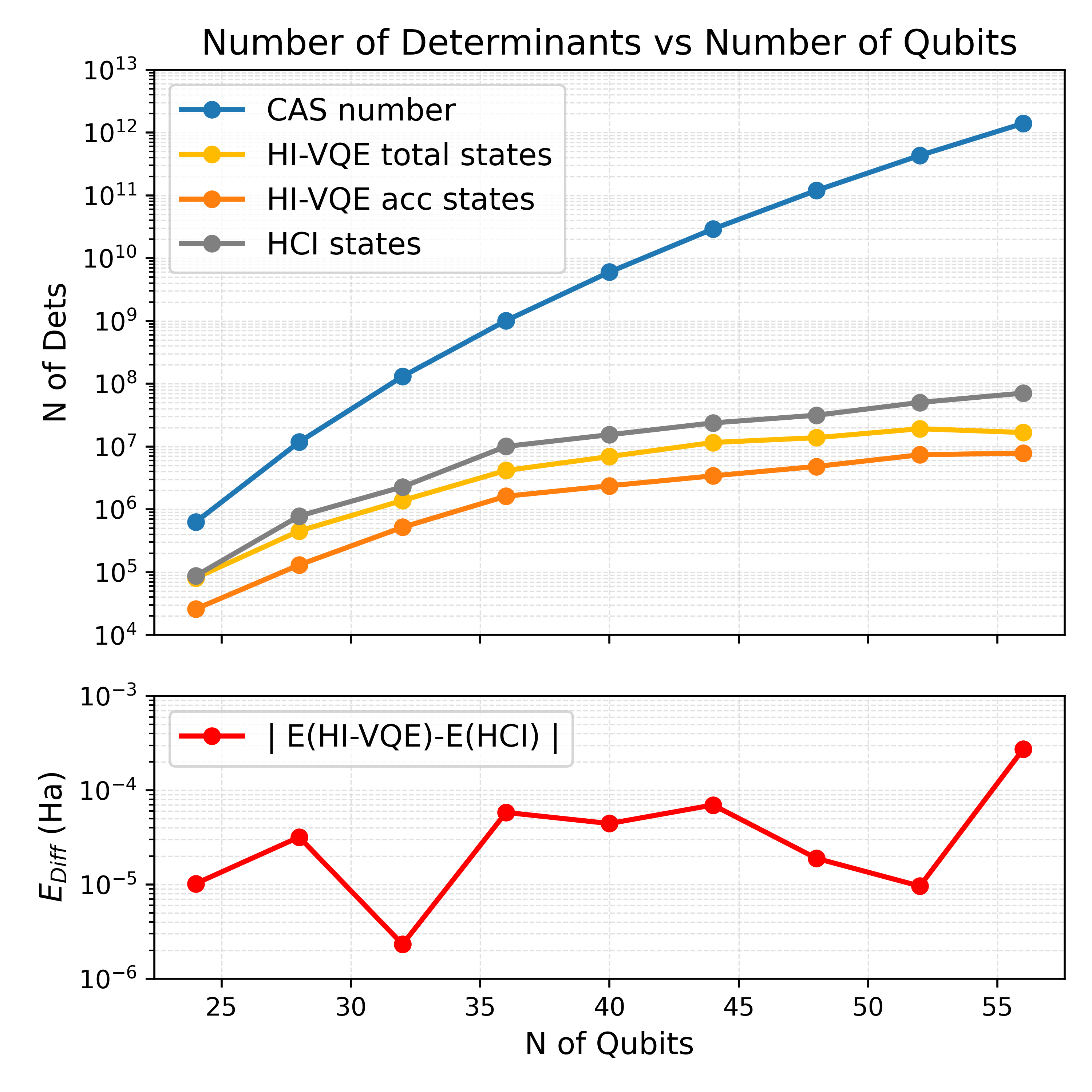}        
    \caption{Top: Determinant scaling with qubit number comparison between CASCI, HI-VQE, and HCI methods for $N_2$ cc-pVDZ at bond distance 3.0 \AA  . Bottom: Absolute energy difference between HI-VQE and HCI.}
    \label{fig:N2_dets_vs_qubits}
\end{figure}

The Complete Active Space (CAS) determinant count (CAS number) grows combinatorially and rapidly becomes intractable as the number of spin orbitals (or qubits) increases ($\binom{28}{7} = 1.4\times10^{12}$), as shown in Figure~\ref{fig:N2_dets_vs_qubits}. To investigate the scaling in a controlled setting, we examined the dissociated $\mathrm{N}_{2}$ molecule at $R = 3.0 \text{\AA}$, varying the number of spin orbitals while keeping the number of electrons fixed (14 electrons). Classical HCI constructs its selected space by expanding a hierarchical excitation tree from a single reference determinant, which leads to a steep increase in the number of selected determinants as the system size grows. However, HI-VQE maintains a substantially more compact determinant representation: both the total QPU-sampled configurations and the extracted configurations from QPU-sampled configurations with amplitude screening remain one to two orders of magnitude smaller than HCI across the full qubit range. It is important to note that the HCI results shown here were carefully converged to achieve energies comparable to those of HI-VQE. These observations demonstrate that QPU sampling can efficiently identify high-weight determinants without relying on exhaustive tree search, enabling scalable many-body approximations at qubit sizes where classical SCI approaches its computational limits. Moreover, because the diagonalization cost scales as $O(N^{3})$ with respect to the number of configurations $N$, the compact determinant space generated by HI-VQE translates directly into significantly more efficient computation within the quantum--classical hybrid framework.

The robustness of HI-VQE in this challenging regime highlights its potential as a practical quantum-classical hybrid algorithm for strongly correlated molecular systems. As \ch{N2} dissociation represents a prototypical test of multireference correlation, the ability of HI-VQE to accurately reproduce HCI-quality results in both equilibrium and dissociated limits demonstrates its suitability for more complex systems, such as transition metal clusters and catalytic reaction paths.

\subsection{[4Fe-4S] cluster}

Beyond the [2Fe–2S] cluster, cubane-type [4Fe–4S] clusters constitute a second prototypical Fe–S motif with even richer multireference structure and a denser manifold of low-lying spin states. In contrast to the dimer, which is often adequately described by a dominant antiferromagnetic coupling between two high-spin Fe(III) centers, the [4Fe–4S] core combines four exchange-coupled high-spin Fe sites arranged in a nearly tetrahedral geometry, leading to extensive spin frustration and competing superexchange pathways. As a result, the ground and thermally accessible excited states involve strong mixing of multiple charge- and spin-localized configurations, including Fe(II)/Fe(III) valence tautomerism and Fe$(d-d)$ excitations that invalidate simple Heisenberg–double-exchange pictures and demand a genuinely multiconfigurational treatment.

From a computational point of view, these features place the [4Fe–4S] cluster firmly in the regime where single-reference methods such as HF, CCSD, and CCSD(T) are qualitatively unreliable, even when combined with large basis sets and scalar-relativistic corrections. While DMRG and SHCI can reach near-CASCI/FCI quality in carefully optimized active spaces of 30–60 electrons in 30–40 orbitals, the corresponding determinant or renormalized-bond dimensions required to converge the low-lying spin ladder typically reach $10^8$ – $10^9$ determinants or very large MPS bond dimensions, pushing the limits of classical resources for realistic models.\cite{sharma2014natchem,lee2023quantumadvantage,Mejuto-Zaera2022FeSCubanes} Recent high-level studies have shown that even for minimal [4Fe–4S] and related model complexes, underscoring the need for explicit many-electron treatments. These observations make [4Fe–4S] clusters a stringent and chemically meaningful benchmark for testing whether HI‑VQE can recover DMRG/SHCI-level energetics while retaining the determinant compactness already demonstrated for [2Fe–2S].

In the following, we therefore extend our HI‑VQE framework to a [4Fe–4S] model system, employing an active space constructed to capture the essential Fe 3d and S 3p degrees of freedom and their associated superexchange pathways. We compare the resulting HI‑VQE ground-state energies to classical references obtained from HF, CCSD, and high-accuracy SHCI/DMRG calculations\cite{li2017spinprojected}, and we analyze both the absolute energy errors and the size and structure of the selected determinant subspaces. This enables us to assess whether quantum-informed sampling continues to yield a more compact but physically meaningful representation of the multireference wavefunction in a regime where classical tree-search-based SCI and tensor-network approaches become particularly demanding.

\begin{figure}[ht]
    \centering
    \includegraphics[width=0.75\linewidth]{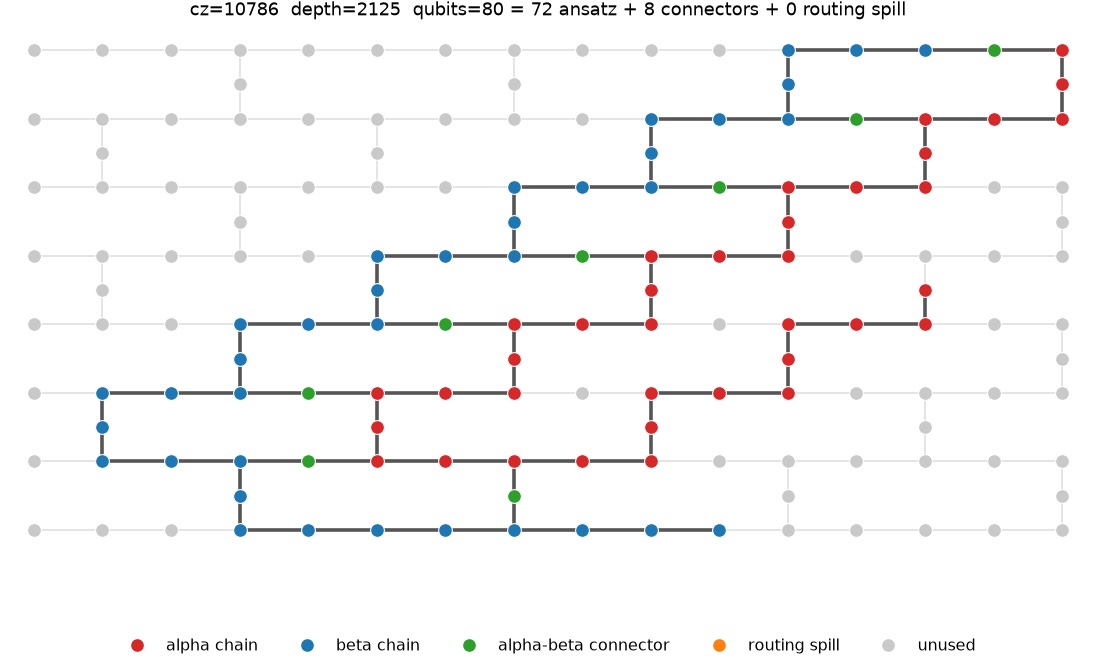}
    \caption{LUCJ ansatz (36 orbitals, \texttt{n\_reps}~$=2$) pinned onto
    \texttt{ibm\_kobe} (156-qubit Heron~r2) by the zigzag embedding. Red/blue: the $\alpha$ and $\beta$ orbital chains; green: $\alpha$--$\beta$ connectors; grey: unused device qubits. 80 physical qubits $=$ 72 ansatz $+$ 8 connectors $+$ 0 routing spill; \texttt{cz}~$=10{,}786$, depth~$=2{,}125$. The identical transpiled template is reused for every iteration generated at the beginning of HI-VQE calculation.}
    \label{fig:fe4s4_lucj_mapping}
\end{figure}

We applied HI-VQE to the $[\mathrm{Fe_4S_4(SCH_3)_4}]^{2-}$ cluster in the TZP-DHK basis using a (54e,\,36o) active space, corresponding to 72 qubits. In contrast to the [2Fe--2S] runs, whose quantum samples were drawn from the \texttt{ibm\_kobe} backend, the substantially larger 72-qubit workload was executed on the supercomputer Fugaku (Fujitsu A64FX): configurations were generated with the sampler from a linear-depth unitary cluster-Jastrow (LUCJ) circuit using zigzag mapping to the physical quibts fig \ref{fig:fe4s4_lucj_mapping}, and each handover diagonalization was carried out with a Qunonva in-house sparse diagonalization solver. The calculation comprised 32 iterations organised in two sampling regimes: an initial phase (iterations 1--18; 12{,}000 shots, variational threshold $\epsilon_{\mathrm{var}}=10^{-6}$) followed by a refinement phase (iterations 19--27; 14{,}000 shots, $\epsilon_{\mathrm{var}}=5\times10^{-7}$) in which the sampling statistics and selection threshold were tightened to accelerate convergence.

The variational energy decreased monotonically over the handover sequence, from $-326.7232736$~Ha at the first iteration to a best value of $-327.1757326$~Ha (iteration 32) for variational energy. Table~\ref{tab:fe4s4} places this result against the classical reference methods. Measured from the near-exact DMRG benchmark ($-327.23964$~Ha), HI-VQE lies $64$~mHa above the reference, whereas the single-reference and truncated selected-CI methods deviate by $402$--$692$~mHa. Expressed as a fraction of the DMRG correlation energy (relative to RHF), HI-VQE recovers $90.77\%$ of the correlation, compared with $41.9\%$ for CCSD, $35.6\%$ for the best available HCI expansion, and $28.1\%$ for CISD. HI-VQE therefore captures more than twice the correlation energy of the coupled-cluster and selected-CI baselines, consistent with the strongly multireference character of the cubane core and the failure of single-reference approaches noted above.

This accuracy is achieved with a compact variational determinant space: the accumulated (union) determinant count grew to $1.27\times10^{8}$ ($2.25\times10^{8}$) determinants reaching the lowest variational energy of this run, which is more than six orders of magnitude below the full $(54e,36o)$ Hilbert space ($\sim\!9.15\times10^{14}$). Reaching this regime, however, is computationally demanding: the tensor-product-expanded sample subspace diagonalized at each iteration reached $9.9\times10^{8}$ states, the associated sparse Hamiltonians contained up to $6.2\times10^{11}$ (sample space) and $5.4\times10^{11}$ (union space) non-zero elements, and peak memory reached $23.9$~GB per rank. The complete route consumed roughly $66,249$~ node-hours on 1{,}920 Fugaku nodes. A complete per-iteration record of the route -- energies, node counts, cost, timings, and job provenance -- is given in Table~\ref{tab:fe4s4_route_cost}, and the corresponding sparse-Hamiltonian structure of the two subspaces diagonalized at each step is summarized in Table~\ref{tab:fe4s4_route_ham}. We emphasise that, unlike the [2Fe--2S] and N$_2$ benchmarks, the $[4\mathrm{Fe}\text{--}4\mathrm{S}]$ run does not yet reach chemical accuracy relative to DMRG; the residual $\sim\!64$~mHa gap in variational-only and $\sim\!46$~mHa gap in variational followed by perturbartive correction reflect the extreme spin frustration and dense low-lying manifold of the cubane cluster. Nonetheless, the demonstration that quantum-informed sampling recovers the large majority of the correlation energy at the 72-qubit scale, while retaining a determinant subspace many orders of magnitude smaller than the formal active space, establishes HI-VQE as a viable route toward this frontier class of strongly correlated systems.

\begin{table}[htbp]
\centering
\caption{$[4\mathrm{Fe}\text{--}4\mathrm{S}]$ cluster
$[\mathrm{Fe_4S_4(SCH_3)_4}]^{2-}$ in the TZP-DHK basis: ground-state energy,
energy error, and number of the variational space determinants for Hartree--Fock, CISD, CCSD, HCI,
and HI-VQE, compared to the DMRG reference.\ Classical reference energies from
Ref.~[35]. HI-VQE energy is the best variational value (iteration 32);
the determinant count is the union subspace size at that iteration.}
\label{tab:fe4s4}
\begin{tabular}{lrrr}
\hline
Method & $E_{\mathrm{total}}$ (Ha) & $\Delta E^{a}$ (mHa) & No.\ of Dets \\
\hline
Hartree--Fock        & $-326.547415$ & $692.2$ & $1$ \\
CISD                 & $-326.742123$ & $497.5$ & -- \\
CCSD                 & $-326.837633$ & $402.0$ & -- \\
HI-VQE(var)          & $-327.175695$ & $63.94$  & 225{,}042{,}563  \\
HI-VQE(var+pt2)      & $-327.192762$ & $47.16$  & 225{,}042{,}563  \\
DMRG                 & $-327.239637$ & $0.0$   & -- \\

\hline
\end{tabular}

\vspace{2pt}
{\footnotesize $^{a}\,E_{m}-E_{\mathrm{DMRG}}$.}
\end{table}

\begin{table}[htbp]
\centering
\footnotesize
\setlength{\tabcolsep}{6pt}
\caption{Per-iteration record of the [4Fe--4S] HI-VQE route (54e,\,36o; 72
qubits for spin-orbitals). Node-hours
are classical solver cost, $(\text{wall}-\text{QPU})\times1920/3600$ , the
part of an iteration that actually consumes nodes for compute; QPU sampling
time is excluded because pricing a $\sim$10~s sampler call at the held node
count measures a reservation choice rather than a physical cost. This case study utilized \texttt{ibm\_kobe} for 16{,}000 shots per iteration. Energies in Hartree; \emph{union dets} is the
spin-symmetrized space actually diagonalized. Correlation recovery is estimated against
the classical reference set from the DMRG work with \emph{this} Hamiltonian (36-orbital /
54-electron:
$E_{\mathrm{DMRG}(D=4000)}=-327.23963690$~Ha, hence
$E_{\mathrm{corr}}=-692.22$~mHa and $\%\,\mathrm{corr} =
(E-E_{\mathrm{RHF}})/(E_{\mathrm{DMRG}}-E_{\mathrm{RHF}})\times100$. The
denominator is DMRG at $D=4000$ and not FCI, so each percentage slightly
\emph{over}estimates the true fraction recovered and should be quoted as ``\%
of the DMRG($D{=}4000$) correlation energy''; the mHa gap is given beside it
because on a 692~mHa denominator the best iteration is simultaneously 90.77\%
and $40\times$ chemical accuracy (1~kcal\,mol$^{-1}=1.594$~mHa).}
\label{tab:fe4s4_route_cost}
\resizebox{\textwidth}{!}{%
\begin{tabular}{rrrrrrl}
\toprule
it & $E$ (Ha) & \% corr & gap (mHa) & union dets & node-h & sampler job \\
\midrule
1 & $-326.72327357$ & 25.40\% & 516.36 & 30{,}114{,}043 & 1{,}228 & \texttt{da5cieeaa69c739kn5jg} \\
2 & $-326.98207035$ & 62.79\% & 257.57 & 34{,}456{,}591 & 1{,}289 & \texttt{da5cocrotlns739bp710} \\
3 & $-326.98194832$ & 62.77\% & 257.69 & 28{,}535{,}520 & 1{,}292 & \texttt{da5dcnk3jnrc73ah62e0} \\
4 & $-327.07317741$ & 75.95\% & 166.46 & 48{,}842{,}228 & 1{,}417 & \texttt{da5e3um1vhnc73flvs90} \\
5 & $-327.10643004$ & 80.76\% & 133.21 & 72{,}122{,}446 & 1{,}633 & \texttt{da5epkjotlns739brj60} \\
6 & $-327.14136061$ & 85.80\% & 98.28 & 108{,}361{,}491 & 1{,}947 & \texttt{da5fcb61vhnc73fm16lg} \\
7 & $-327.15155043$ & 87.27\% & 88.09 & 117{,}202{,}159 & 2{,}032 & \texttt{da5fldmaa69c739kqjng} \\
8 & $-327.15846314$ & 88.27\% & 81.17 & 131{,}762{,}639 & 2{,}199 & \texttt{da5fulmaa69c739kqtc0} \\
9 & $-327.15841750$ & 88.27\% & 81.22 & 120{,}377{,}903 & 1{,}981 & \texttt{da5g8guaa69c739kr7ag} \\
10 & $-327.15841726$ & 88.27\% & 81.22 & 121{,}526{,}889 & 1{,}983 & \texttt{da5gjd6aa69c739kri90} \\
11 & $-327.16088715$ & 88.62\% & 78.75 & 133{,}027{,}782 & 2{,}185 & \texttt{da5grj6aa69c739krr7g} \\
12 & $-327.16153634$ & 88.72\% & 78.10 & 133{,}867{,}767 & 2{,}138 & \texttt{da5h5361vhnc73fm30qg} \\
13 & $-327.16364122$ & 89.02\% & 76.00 & 147{,}269{,}433 & 2{,}311 & \texttt{da5hea43jnrc73aha76g} \\
14 & $-327.16704312$ & 89.51\% & 72.59 & 164{,}318{,}318 & 2{,}516 & \texttt{da5hr7s3jnrc73ahanfg} \\
15 & $-327.16799553$ & 89.65\% & 71.64 & 155{,}738{,}634 & 2{,}387 & \texttt{da5i5quaa69c739kt8q0} \\
16 & $-327.16966805$ & 89.89\% & 69.97 & 164{,}243{,}743 & 2{,}481 & \texttt{da5iggm1vhnc73fm4esg} \\
17 & $-327.16929383$ & 89.84\% & 70.34 & 125{,}362{,}239 & 1{,}998 & \texttt{da5ir8eaa69c739kttng} \\
18 & $-327.16929325$ & 89.84\% & 70.34 & 125{,}327{,}692 & 1{,}997 & \texttt{da5j4njotlns739c04eg} \\
19 & $-327.16929808$ & 89.84\% & 70.34 & 131{,}865{,}977 & 2{,}001 & \texttt{da5jea6aa69c739kuk3g} \\
20 & $-327.17160760$ & 90.17\% & 68.03 & 160{,}267{,}928 & 2{,}451 & \texttt{da5jofuaa69c739kuvi0} \\
21 & $-327.17142593$ & 90.15\% & 68.21 & 134{,}455{,}503 & 2{,}052 & \texttt{da5k2uu1vhnc73fm6660} \\
22 & $-327.17137064$ & 90.14\% & 68.27 & 133{,}099{,}376 & 2{,}023 & \texttt{da5kc4rotlns739c1dt0} \\
23 & $-327.17172290$ & 90.19\% & 67.91 & 144{,}897{,}892 & 2{,}249 & \texttt{da5knebotlns739c1qh0} \\
24 & $-327.17244167$ & 90.29\% & 67.20 & 153{,}017{,}058 & 2{,}364 & \texttt{da5l2hm1vhnc73fm77hg} \\
25 & $-327.17212032$ & 90.25\% & 67.52 & 125{,}979{,}175 & 1{,}926 & \texttt{da6d3jesidac73ae0pjg} \\
26 & $-327.17245314$ & 90.29\% & 67.18 & 134{,}839{,}417 & 1{,}741 & \texttt{da6e5nesidac73ae1vrg} \\
27 & $-327.17247171$ & 90.30\% & 67.17 & 132{,}280{,}585 & 1{,}652 & \texttt{da6f0dm0ukec7381vv10} \\
28 & $-327.17379629$ & 90.49\% & 65.84 & 212{,}236{,}565 & 2{,}461 & \texttt{da6gfr60ukec73821pq0} \\
29 & $-327.17455646$ & 90.60\% & 65.08 & 217{,}321{,}381 & 2{,}466 & \texttt{da6huu6sidac73ae6dpg} \\
30 & $-327.17521283$ & 90.69\% & 64.42 & 218{,}951{,}283 & 2{,}679 & \texttt{da6trfjsq5js73bj7fn0} \\
31 & $-327.17515087$ & 90.68\% & 64.49 & 210{,}963{,}292 & 2{,}534 & \texttt{da6vfousidac73aemqj0} \\
32 & $-327.17573255$ & 90.77\% & 63.90 & 225{,}042{,}563 & 2{,}638 & \texttt{da72k0c6l22c73dn0tb0} \\
\midrule
\multicolumn{5}{l}{\textbf{Total: 66{,}249 node-hours = 34.5 h at 1920 nodes}} \\
\bottomrule
\end{tabular}}
\end{table}

\begin{table}[htbp]
\centering
\footnotesize
\setlength{\tabcolsep}{3pt}
\caption{Per-iteration sparse-Hamiltonian structure for the [4Fe--4S] HI-VQE
route, iterations 1--32. Two distinct Hamiltonians are diagonalized each
iteration, in two separate MPI jobs: the \emph{full sample space}, whose
dimension is the tensor-product-expanded quantum-sample space before
truncation, and the \emph{HI-VQE union} space, the spin-symmetrized
accumulated variational basis that yields the reported eigenvalue. ``NNZ'' is
the global number of non-zero matrix elements and $\langle$nnz/row$\rangle$
the mean non-zeros per row (global NNZ $\div$ dimension). ``truncated core''
is the determinant count the sample space is clamped to before CE, (20{,}000{,}000 determinants up to iteration~27 and 60{,}000{,}000 determinants from iteration~28) to drop the marginally contributing states; ``accum.\
states'' is the pool carried over to the next iteration to preserve important
core determinants. The memory columns (mem/rank) are based on the usage per rank in the Fugaku configuration of 1920
A64FX nodes with one MPI rank per node which accesses the A64FX's 32~GB of HBM2.}
\label{tab:fe4s4_route_ham}
\resizebox{\textwidth}{!}{%
\begin{tabular}{rrrrrrrrrrr}
\toprule
 & \multicolumn{5}{c}{Sample space (tensor-product)} & \multicolumn{5}{c}{Union space (variational dets)} \\
\cmidrule(lr){2-6}\cmidrule(lr){7-11}
it & dim & NNZ & $\langle$nnz/row$\rangle$ & truncated core & mem/rank & dim & NNZ & $\langle$nnz/row$\rangle$ & accum.\ states & mem/rank \\
 & & & & & (GB) & & & & & (GB) \\
\midrule
1 & 987{,}782{,}041 & 122{,}353{,}277{,}293 & 124 & 20{,}000{,}000 & 15.4 & 30{,}114{,}043 & 780{,}904{,}703 & 26 & 67{,}387 & 2.7 \\
2 & 993{,}951{,}729 & 87{,}504{,}540{,}145 & 88 & 20{,}000{,}000 & 14.8 & 34{,}456{,}591 & 26{,}196{,}609{,}893 & 760 & 3{,}571{,}637 & 3.0 \\
3 & 792{,}704{,}025 & 623{,}111{,}837{,}701 & 786 & 20{,}000{,}000 & 23.8 & 28{,}535{,}520 & 10{,}918{,}872{,}956 & 383 & 3{,}571{,}904 & 2.7 \\
4 & 984{,}829{,}924 & 104{,}424{,}909{,}512 & 106 & 20{,}000{,}000 & 15.1 & 48{,}842{,}228 & 76{,}522{,}138{,}684 & 1{,}567 & 10{,}471{,}155 & 3.6 \\
5 & 987{,}844{,}900 & 93{,}480{,}083{,}540 & 95 & 20{,}000{,}000 & 14.9 & 72{,}122{,}446 & 162{,}276{,}199{,}402 & 2{,}250 & 28{,}016{,}203 & 4.6 \\
6 & 971{,}319{,}556 & 110{,}767{,}825{,}444 & 114 & 20{,}000{,}000 & 15.2 & 108{,}361{,}491 & 286{,}530{,}781{,}771 & 2{,}644 & 54{,}638{,}135 & 6.1 \\
7 & 972{,}317{,}124 & 103{,}412{,}968{,}540 & 106 & 20{,}000{,}000 & 15.1 & 117{,}202{,}159 & 320{,}608{,}782{,}787 & 2{,}736 & 73{,}024{,}170 & 6.5 \\
8 & 962{,}426{,}529 & 114{,}844{,}736{,}821 & 119 & 20{,}000{,}000 & 15.3 & 131{,}762{,}639 & 386{,}583{,}713{,}823 & 2{,}934 & 90{,}768{,}549 & 7.3 \\
9 & 857{,}552{,}656 & 604{,}424{,}517{,}184 & 705 & 20{,}000{,}000 & 23.5 & 120{,}377{,}903 & 285{,}273{,}822{,}411 & 2{,}370 & 91{,}042{,}799 & 6.3 \\
10 & 863{,}654{,}544 & 599{,}644{,}890{,}244 & 694 & 20{,}000{,}000 & 23.4 & 121{,}526{,}889 & 286{,}202{,}914{,}719 & 2{,}355 & 91{,}554{,}944 & 6.3 \\
11 & 972{,}816{,}100 & 109{,}786{,}703{,}516 & 113 & 20{,}000{,}000 & 15.2 & 133{,}027{,}782 & 381{,}262{,}386{,}396 & 2{,}866 & 96{,}000{,}408 & 7.2 \\
12 & 954{,}810{,}000 & 163{,}413{,}233{,}776 & 171 & 20{,}000{,}000 & 16.1 & 133{,}867{,}767 & 360{,}801{,}790{,}027 & 2{,}695 & 99{,}371{,}090 & 7.1 \\
13 & 967{,}769{,}881 & 115{,}992{,}782{,}981 & 120 & 20{,}000{,}000 & 15.3 & 147{,}269{,}433 & 430{,}872{,}487{,}229 & 2{,}926 & 100{,}000{,}000 & 7.8 \\
14 & 966{,}215{,}056 & 116{,}108{,}466{,}576 & 120 & 20{,}000{,}000 & 15.3 & 164{,}318{,}318 & 512{,}445{,}725{,}254 & 3{,}119 & 100{,}000{,}000 & 8.7 \\
15 & 972{,}130{,}041 & 107{,}742{,}559{,}309 & 111 & 20{,}000{,}000 & 15.2 & 155{,}738{,}634 & 461{,}597{,}353{,}800 & 2{,}964 & 100{,}000{,}000 & 8.2 \\
16 & 963{,}729{,}936 & 116{,}344{,}864{,}756 & 121 & 20{,}000{,}000 & 15.3 & 164{,}243{,}743 & 498{,}613{,}255{,}457 & 3{,}036 & 100{,}000{,}000 & 8.6 \\
17 & 855{,}270{,}025 & 596{,}767{,}252{,}409 & 698 & 20{,}000{,}000 & 23.4 & 125{,}362{,}239 & 292{,}239{,}323{,}883 & 2{,}331 & 100{,}000{,}000 & 6.4 \\
18 & 853{,}866{,}841 & 612{,}268{,}320{,}429 & 717 & 20{,}000{,}000 & 23.7 & 125{,}327{,}692 & 291{,}455{,}079{,}662 & 2{,}326 & 100{,}000{,}000 & 6.4 \\
19 & 862{,}773{,}129 & 602{,}579{,}168{,}025 & 698 & 20{,}000{,}000 & 23.5 & 131{,}865{,}977 & 293{,}414{,}059{,}107 & 2{,}225 & 100{,}000{,}000 & 6.5 \\
20 & 959{,}326{,}729 & 119{,}656{,}112{,}297 & 125 & 20{,}000{,}000 & 15.4 & 160{,}267{,}928 & 486{,}435{,}346{,}700 & 3{,}035 & 100{,}000{,}000 & 8.5 \\
21 & 851{,}005{,}584 & 616{,}618{,}098{,}772 & 725 & 20{,}000{,}000 & 23.7 & 134{,}455{,}503 & 313{,}137{,}582{,}513 & 2{,}329 & 100{,}000{,}000 & 6.7 \\
22 & 852{,}231{,}249 & 624{,}788{,}079{,}997 & 733 & 20{,}000{,}000 & 23.9 & 133{,}099{,}376 & 301{,}336{,}809{,}970 & 2{,}264 & 100{,}000{,}000 & 6.6 \\
23 & 979{,}126{,}681 & 105{,}601{,}661{,}317 & 108 & 20{,}000{,}000 & 15.1 & 144{,}897{,}892 & 406{,}565{,}574{,}026 & 2{,}806 & 100{,}000{,}000 & 7.6 \\
24 & 965{,}282{,}761 & 111{,}962{,}339{,}749 & 116 & 20{,}000{,}000 & 15.2 & 153{,}017{,}058 & 452{,}307{,}837{,}544 & 2{,}956 & 100{,}000{,}000 & 8.1 \\
25 & 861{,}892{,}164 & 573{,}218{,}271{,}136 & 665 & 20{,}000{,}000 & 23.0 & 125{,}979{,}175 & 289{,}596{,}746{,}179 & 2{,}299 & 100{,}000{,}000 & 6.9 \\
26 & 968{,}080{,}996 & 136{,}174{,}144{,}152 & 141 & 20{,}000{,}000 & 15.6 & 134{,}839{,}417 & 357{,}668{,}468{,}961 & 2{,}653 & 100{,}000{,}000 & 6.6 \\
27 & 950{,}797{,}225 & 201{,}835{,}539{,}829 & 212 & 20{,}000{,}000 & 16.9 & 132{,}280{,}585 & 334{,}831{,}578{,}137 & 2{,}531 & 100{,}000{,}000 & 6.3 \\
28 & 967{,}832{,}100 & 110{,}718{,}151{,}416 & 114 & 60{,}000{,}000 & 15.3 & 212{,}236{,}565 & 478{,}599{,}799{,}793 & 2{,}255 & 110{,}728{,}918 & 8.8 \\
29 & 972{,}379{,}489 & 99{,}639{,}287{,}177 & 102 & 60{,}000{,}000 & 14.8 & 217{,}321{,}381 & 503{,}991{,}868{,}069 & 2{,}319 & 115{,}176{,}243 & 9.4 \\
30 & 972{,}254{,}761 & 103{,}616{,}493{,}029 & 107 & 60{,}000{,}000 & 14.9 & 218{,}951{,}283 & 530{,}642{,}062{,}585 & 2{,}424 & 123{,}315{,}212 & 9.5 \\
31 & 872{,}670{,}681 & 557{,}217{,}848{,}997 & 639 & 60{,}000{,}000 & 22.7 & 210{,}963{,}292 & 416{,}497{,}418{,}448 & 1{,}974 & 123{,}330{,}412 & 8.9 \\
32 & 972{,}067{,}684 & 102{,}561{,}121{,}588 & 106 & 60{,}000{,}000 & 15.0 & 225{,}042{,}563 & 541{,}444{,}181{,}039 & 2{,}406 & 126{,}563{,}465 & 9.7 \\
\midrule
max & 993{,}951{,}729 & 624{,}788{,}079{,}997 & 786 & 60{,}000{,}000 & 23.9 & 225{,}042{,}563 & 541{,}444{,}181{,}039 & 3{,}119 & 126{,}563{,}465 & 9.7 \\
\bottomrule
\end{tabular}}
\end{table}

\section{Discussion}

\subsection{Path dependence of the cumulative subspace}
\label{sec:path-dependence}

The HI-VQE workflow advances through two deliberately separated tracks. At each iteration, the Hamiltonian is diagonalized in the subspace spanned by the configurations sampled from the QPU in that iteration alone; the resulting eigenvector is amplitude-screened and used to parameterize the circuit for the next iteration, so that a lower energy obtained from the circuit-sampled subspace defines the preferred search direction. In parallel, all sampled configurations are accumulated into a growing total subspace, which supplies the reported variational energy but is withheld from the circuit parameterization so that it cannot destabilize the optimization.

This separation stabilizes energy convergence and prevents abrupt subspace fluctuations, but it also renders the trajectory path dependent. The search direction at every iteration is fixed by the most recent circuit-sampled subspace, which in turn descends from the sampling of the first iteration; because the accumulated subspace never feeds back into the parameterization, no mechanism exists to redirect the circuit toward regions of the configuration space that earlier iterations failed to reach. Since the Hamiltonian couples a given determinant only to those differing from it by at most a single or double excitation, each iteration expands the sampled region locally around its current seed and the quantum circuit property rather than exploring the active space independently.

The relevance of this path dependence is specific to the system. Iron--sulfur clusters support a dense manifold of low-lying states that differ in how the local spins of the Fe centers are coupled, and wavefunction optimizations carried out in a restricted variational manifold are known to exhibit multiple local minima associated with these distinct spin couplings~\cite{sharma2014natchem, li2017spinprojected, li2019pcluster, limanni2021cubanes}. In established treatments the minimum that is reached is controlled deliberately: either by enumerating broken-symmetry references and converging one solution per basin, or by adapting or projecting the wavefunction onto a definite spin sector. The present workflow imposes no such control. The ansatz is not spin-adapted, the sampled configurations are not projected onto a target spin coupling, and no broken-symmetry reference is enforced at any stage. Consequently, which basin a trajectory settles into is determined by the configurations that happen to be sampled in the first iteration rather than by any property of the method, and independent runs of the same calculation may resolve different members of this near-degenerate manifold.

This is consistent with the independent [2Fe--2S] (30e,20o) runs reported in Sec. \ref{[2Fe-2S]_cluster}. The runs converge to comparable total energies, with a spread of $\sim$0.2~mHa, yet the determinant sets they retain overlap only weakly---the pattern expected when nearly degenerate solutions of different spin-coupling character are reached from
different starting points. We note that weakly overlapping supports alone do not establish that the runs occupy physically distinct basins, since a single state also admits many nearly equivalent compact representations. Separating the two requires comparing the converged runs in terms of $\langle \hat{S}^2\rangle$, local Fe spins, the spin--spin correlation $\langle \vec{S}_{\rm Fe1}\cdot \vec{S}_{\rm Fe2} \rangle$, and orbital occupation numbers. That comparison, together with a quantitative characterization of the overlap---mean pairwise Jaccard index of the retained determinant sets, the amplitude-weighted overlap $|\langle \Psi_i | \Psi_j \rangle|$, and the dependence of both on the amplitude cutoff used to define the support---is deferred to a forthcoming analysis.

A direct consequence of this path dependence is that the union of determinants collected from independent runs spans a larger portion of the relevant space than any single trajectory, and lowers the variational energy by some extent, relative to the best single run. Because a union is variationally superior to its members by construction, this figure alone does not establish that independent trajectories carry complementary information; the appropriate control is a single trajectory grown to the same subspace dimension $D_{\rm union}$. That control experiment is in progress and will be reported alongside the overlap analysis above.


\subsection{Comparison to selected-CI-family and quantum-sampling methods}

The basin-trapping phenomenon described above is not unique to HI-VQE; it is a structural feature of any iterative selected-CI or quantum-sampling method that expands its variational space from a previously converged state. Classical SHCI and HCI similarly build their determinant spaces by local expansion from a seed configuration, typically Hartree--Fock, and are therefore susceptible to the same trapping when the wavefunction landscape contains multiple near-degenerate basins separated by regions of negligible Hamiltonian connectivity.
It is worth noting that stochastic selected-CI methods such as Trim-CI are particularly susceptible to this multi-basin trapping: because their determinant selection involves stochastic sampling, individual runs may converge to different local basins, and achieving a reliably low energy may require aggregating results from several thousand independent runs. This stochastic variance compounds the basin-trapping effect, making it difficult to ascertain whether a given Trim-CI energy reflects the global minimum or merely the best basin encountered among a limited number of trials.

The key distinction is that HI-VQE, by seeding each iteration with QPU-sampled configurations, has a mechanism to reach distant regions of Hilbert space that are inaccessible through local single- and double-excitation expansion alone (Figure~\ref{fig:Subspace_expansion}). This is precisely why HI-VQE achieves comparable or better energies with one to two orders of magnitude fewer determinants than HCI. The quantum processor acts not as a more accurate energy estimator, but as a \emph{configuration explorer} whose sampling distribution is shaped by entanglement rather than by excitation-rank connectivity. The practical consequence is that HI-VQE discovers high-weight determinants that classical tree-search methods reach only at much greater computational expense.

This interpretation also clarifies the relationship between HI-VQE and IBM's sample-based quantum diagonalization (SQD)~\cite{robledo2025quantumcentric}. Both methods use the quantum processor to propose configurations and a classical solver to diagonalize. The critical difference is iterativity: SQD samples a single quantum state and post-processes classically, whereas HI-VQE feeds the diagonalization result back into the circuit optimization, refining the sampling distribution across iterations, which balances the QPU and CPU resources. On the [2Fe--2S] cluster, this iterative refinement yields an order-of-magnitude improvement in energy error (0.83~mHa vs.\ 151.9~mHa) at comparable subspace size ($\sim$63M vs.\ $\sim$56M determinants), directly demonstrating that iterative quantum-classical feedback is substantially more sample-efficient than single-shot sampling for strongly correlated systems.

\subsection{Calculation result of [4Fe--4S] cluster}

The residual 63.9~mHa gap between HI-VQE and the DMRG reference on [4Fe--4S] (Table~\ref{tab:fe4s4}) deserves careful contextualization. This gap does \emph{not} imply that HI-VQE has failed on this system; rather, it reflects the extreme difficulty of the (54e,36o) active space, which hosts a CAS dimension of $9.15 \times 10^{14}$ and a dense manifold of near-degenerate spin states arising from four exchange-coupled Fe centers. In this context, the relevant metric is not the absolute deviation from DMRG but the fraction of correlation energy recovered. HI-VQE captures 90.8\% of the total correlation energy (relative to RHF), compared with 41.9\% for CCSD and 28.1\% for CISD. HI-VQE therefore recovers more than twice the correlation energy of the best available coupled-cluster and truncated CI expansions, consistent with the strongly multireference character of the cubane core.

\subsection{Required computation resource}

We disclosed the resource usage of one case of 4Fe--4S calculation in Tables~\ref{tab:fe4s4_route_cost} and~\ref{tab:fe4s4_route_ham},about 66,000 node-hours on 1,920 Fugaku nodes up to 32 iterations until convergence. Resource accounting of this kind is largely absent from the hybrid quantum-classical algorithm literature, making it difficult to assess actual costs, identify bottlenecks, or plan future deployments. We therefore regard the detailed per-iteration record of energies, node counts, wall times, sparse-Hamiltonian dimensions, and peak memory footprints (Tables~\ref{tab:fe4s4_route_cost}--\ref{tab:fe4s4_route_ham}) as a contribution in its own right.

Several features of this accounting deserve emphasis. First, the tensor-product-expanded sample subspace diagonalized at each iteration reached up to $\sim$$10^{9}$ states generated from only QPU sampled states to ensure the connectivity between remote excitations from Ansatz, with associated sparse Hamiltonians containing up to $6.2 \times 10^{11}$ non-zero elements (Table~\ref{tab:fe4s4_route_ham}). These dimensions are far beyond what a single workstation or even a mid-scale HPC cluster can handle, confirming that the [4Fe--4S] benchmark genuinely requires supercomputer-class classical resources. Second, the per-rank peak memory and the Qunova Sparse CI eigensolver times of up to $\sim$1,400~s per iteration identify sparse-matrix diagonalization as the dominant classical bottleneck, not Hamiltonian construction or parameter optimization.

A methodologically grounded comparison can be drawn against TrimCI, a classical selected-CI method that, like HI-VQE, constructs the ground-state wavefunction through determinant-state discovery. We estimated the expected cost-to-solution of TrimCI on the same (54e,36o) active space by executing 995 independent Phase-0 runs with different random seeds on single Fugaku node per calculation. All runs were terminated by out-of-memory with the 28\,GB memory limit before reaching the determinant target of $6.18 \times 10^{7}$; extrapolating from the median anchor point (325,138 determinants reached in 12,217\,s per run) with the measured local growth exponent $b = 1.34$ yields an estimated batch cost of $\sim$245,000 node-hours for 64 independent runs, the minimum ensemble size needed to locate a competitive basin to reach a variational energy of $-327.1747$\,Ha. HI-VQE achieves an essentially identical energy ($-327.1746$\,Ha at iteration~29) at a cumulative cost of $\sim$58,000 node-hours. The key distinction here is not only speed but \emph{consistency}: whereas TrimCI is stochastic and requires many independent random initializations to locate energetically favorable determinant seed space, with no guarantee that any single run will reach the global minimum, HI-VQE follows a deterministic, iterative route through quantum-guided sampling that converges reproducibly to a competitive energy. This reproducibility is a practical advantage in its own right, as it allows reliable estimation of the resources required for a given accuracy target without the uncertainty inherent in stochastic ensemble approaches. As a further reference point, the SQD method reported an energy of $-326.913$\,Ha on the same [4Fe--4S] system\cite{shirakawa2025closedloopcalculationselectronicstructure}; HI-VQE surpasses this energy ($-326.982$\,Ha) by the second iteration, illustrating that iterative quantum-classical feedback can reach lower energies with substantially fewer resources than single-shot quantum sampling. Together, these comparisons suggest that the quantum-guided determinant discovery strategy employed by HI-VQE offers a promising and resource-efficient pathway for selected-CI-quality calculations on strongly correlated systems at this scale.

These numbers also establish a baseline for projecting future costs. The [4Fe--4S] system uses 72 qubits and a (54e,36o) active space; biologically relevant targets such as the full FeMo cofactor (FeMoco) of nitrogenase would require $\sim$100--150 qubits and active spaces of (100+e, 60--80o). Extrapolating from the observed scaling, in particular, the $O(N^3)$ dependence of diagonalization cost on subspace dimension and the approximately linear growth of the union subspace with iterations, suggests that FeMoco-scale HI-VQE would require $\sim$$10^6$--$10^7$ node-hours and correspondingly larger quantum sampling budgets, placing it within reach of exascale classical resources paired with next-generation quantum processors.

\subsection{The role of the quantum processor}

The results presented here suggest a reframing of the near-term role of quantum hardware in chemistry. In HI-VQE the quantum processor never estimates an energy; its contribution is \emph{configuration discovery}---proposing
high-weight Slater determinants for a classical solver to diagonalize. This changes what is demanded of the device: not accurate expectation values, which degrade rapidly under noise, but a sampling distribution whose support overlaps the determinants that matter. The reported energy remains a variational upper bound whatever the sampling quality, so noise costs efficiency rather than correctness.

Whether such sampling offers a genuine advantage over classical selection heuristics is an open question, and one that has been argued in the negative: for the systems examined so far, classical heuristics have been reported to generate more compact expansions than quantum sampling does\cite{Reinholdt2025}. That comparison concerns compactness; the two approaches also differ in what they can reach. Classical selection is driven by the magnitude of the coupling $H_{ij}$, which by the Slater--Condon rules vanishes beyond a double excitation\cite{huron1973iterative,holmes2016heat}, so each iteration admits only the singles-and-doubles neighborhood of the determinants already retained, and high-rank determinants arrive only after several rounds of combinatorially growing intermediate sets. Quantum sampling draws from $|\langle x|\Psi(\boldsymbol{\theta})\rangle|^2$, fixed by the prepared state rather than by its coupling to the current expansion, so determinants of arbitrary rank carry weight directly. The reachable set is a property of the circuit rather than of the two-body structure of the Hamiltonian, and is adjustable by design. This leaves a measurable trace: since $H_{ij}$ vanishes beyond a double excitation, the number of non-zero couplings within the sampled subspace falls as the sampled configurations become mutually more remote in excitation rank. At a given subspace dimension, the sparsity of the subspace Hamiltonian is therefore an indirect diagnostic of how far the sampling reaches---a low non-zero density indicating that much of what the device proposed could not have been connected to the rest by a single step of classical expansion.

What this buys is controllability, not coverage: a circuit whose support misses the important region explores no better than the classical neighborhood. Nor does it displace classical selection---HI-VQE composes the two. Tensor product (TP) treats the $\alpha$ and $\beta$ halves of each sampled bitstring as independent and spans the Cartesian product of the distinct spin-strings, so $N_\alpha$ and $N_\beta$ observed strings define $N_\alpha\times N_\beta$ determinants, most never observed individually. Classical expansion (CE) generates single and double excitations from the highest-amplitude configurations, where the Slater--Condon couplings are exact and cheap to screen\cite{pellowjarman2025hivqe}. Sampling then need not pay for what deterministic construction does well: the circuit supplies the non-local configurations, and TP and CE close the neighborhood around them. The cost should be stated plainly---a product of spin-string sets is a rectangular subspace, whereas the support of a CI wavefunction is not a product set, so TP buys coverage at the expense of compactness. This is part of what compactness comparisons measure\cite{Reinholdt2025}, and it is what the amplitude screening in the handover step controls.

\subsection{Implications for validation and outlook}

Our results do not settle the efficiency question, and they do not yet characterize the sampled subspaces themselves; the overlap analysis that would do so is deferred to the forthcoming work described in
Sec.~\ref{sec:path-dependence}. A structural point stands independently of it. Iron--sulfur clusters support near-degenerate solutions differing in how the local Fe spins are coupled, and the present workflow imposes no constraint that selects among them---the ansatz is not spin-adapted and no broken-symmetry reference is enforced---so the solution a run reaches is fixed by its first iteration rather than by the method. Any separation between such solutions would belong to the restricted variational manifold at a fixed subspace size, not to the Hamiltonian. The risk is then not that correlation energy is hidden elsewhere, but that a calculation converges cleanly by its own diagnostics while representing one member of a near-degenerate manifold. SHCI with its perturbative correction and DMRG at increasing bond dimension approach the same limit in principle; at the subspace sizes affordable for {[4Fe--4S]} they may not. We therefore suggest reporting convergence across the \emph{union} of independent runs---measured against a single run grown to the same dimension---as a complementary criterion, and identify the determinant-overlap and spin-observable measurements that would establish whether it is required.

Finally, iron--sulfur clusters are not merely computational benchmarks; they are the functional cores of nitrogenase, ferredoxins and respiratory-chain complexes whose mechanisms remain incompletely understood at the electronic-structure level~\cite{beinert1997science,sharma2014natchem,zhang2022oxidmed,zhai2023nitrogenase}. Recovering 90.8\% of the correlation energy of {[4Fe--4S]} in a subspace orders of magnitude below the formal CAS dimension falls well short of chemical accuracy; we present it as evidence of feasibility at this scale rather than of parity with established multireference treatments, DMRG having already resolved the low-lying spin manifold of these clusters in detail\cite{sharma2014natchem}. Extending the workflow to {FeMoco} will require progress on subspace compactness and on the sampling efficiency questioned above; we regard it as a defined target rather than an immediate one.

\section{Conclusion} 

This study demonstrates that the HI-VQE can reliably extend quantum-classical hybrid simulation to the frontiers of strong electronic correlation. By benchmarking against state-of-the-art classical approaches, we showed that HI-VQE reproduces HCI-level energies for both the nitrogen molecule and the [2Fe–2S] cluster within millihartree accuracy. HI-VQE achieves chemical-accuracy-level agreement with classical HCI across the entire \ch{N2} dissociation coordinate while requiring substantially fewer determinants and reduced computational resources than HCI, demonstrating that quantum-informed subspace expansion can outperform purely classical determinant selection even for multireference systems of realistic scale.

Beyond these benchmarks, the results establish HI-VQE as a practical framework for quantum-enhanced electronic-structure theory. Its iterative handover between quantum sampling and classical diagonalization provides a scalable route to treat localized and delocalized correlation on equal footing. As quantum processors mature, this approach offers a compelling path toward first-principles modeling of bioinorganic cofactors, catalytic reaction networks, and correlated materials that currently challenge classical computation. The present work thus marks a step toward the routine use of quantum resources in chemically and biologically meaningful simulations.

\section{Acknowledgement}

We acknowledge the RIKEN Center for Computational Science (R-CCS) for providing access to the supercomputer Fugaku and the IBM Quantum System Two (\texttt{ibm\_kobe} and \texttt{ibm\_fez}) through the JHPC-quantum Test User Program. We thank IBM for providing quantum computing credits through the IBM Quantum Startup Program, which enabled the quantum computations presented in this work.

\bibliography{references_Fe-Sclusters}

\end{document}